\documentclass[pre,aps,showpacs,12pt,a4paper]{revtex4}

\usepackage[dvips]{graphics}
\usepackage[dvips]{graphicx}
\usepackage{float}
\usepackage{epsfig}
\usepackage{tabularx,longtable}
\usepackage{color}
\usepackage{fancybox}
\usepackage{amsmath}
\usepackage{ulem}

\newcommand \be {\begin{equation}}
\newcommand \ee {\end{equation}}
\newcommand \beq {\begin{eqnarray}}
\newcommand \eeq {\end{eqnarray}}

 \def\(({\left(}
 \def\)){\right)}

\def \nn{\nonumber}
\def \beqna{\begin{eqnarray}}
\def \eeqna{\end{eqnarray}}
\def \beq{\begin{equation}}
\def \eeq{\end{equation}}
\def \ln{{\rm ln}}

\def \ln{{\rm ln}}

\def \ab2{\alpha\beta^2}

\newcommand \bea {\begin{eqnarray} \nonumber }
\newcommand \eea {\end{eqnarray}}
\newcommand \de {\delta}

\begin{document}

\title{Negative fluctuation-dissipation ratios in the backgammon model}
\author{A.Garriga$^{1,4}$, I.Pagonabarraga$^2$ and F.Ritort$^{2,3}$}
\affiliation{$^1$ Fundaci\'o Centre Piti\'us d'Estudis Avan\c cats,
Palau de Congressos, 07840, Sta. Eul\`aria, Ibiza
(Spain).}\affiliation{$^2$ Departament de F\'{\i}sica Fonamental,
Facultat de F\'{\i}sica, Universitat de Barcelona, Diagonal 647,
08028 Barcelona (Spain).} \affiliation{$^3$ CIBER-BBN, Networking
Centre on Bioengineering, Biomaterials and
Nanomedicine}\affiliation{$^4$ Departament de Tecnologies de la
Informaci\'o i les Comunicacions, Universitat Pompeu Fabra, Passeig
de Circumval.laci\'o 8, 08003, Barcelona (Spain).}

\date{\today}

\begin{abstract}
We analyze fluctuation-dissipation relations in the Backgammon
model: a system that displays glassy behavior at zero temperature
due to the existence of entropy barriers. We study local and global
fluctuation relations for the different observables in the model.
For the case of a global perturbation we find a unique negative
fluctuation-dissipation ratio that is independent of the observable
and which diverges linearly with the waiting time. This result
suggests that a negative effective temperature can be observed in
glassy systems even in the absence of thermally activated processes.

\end{abstract}

\maketitle

\section{INTRODUCTION}

Understanding nonequilibrium systems remains one of the major
open problems in modern physics. In the last years many theoretical
and experimental studies have focused on the extension of the
concept of temperature to the nonequilibrium regime \cite{jou}.

Glassy systems are adequate for testing nonequilibrium
generalizations of thermodynamic concepts. Glassy materials display
extremely slow dynamics as they approach the amorphous solid phase
from the liquid phase \cite{young}. Below the glass transition
temperature, relaxation times become huge and time-translational
invariance (TTI) is lost meaning that two-time correlation and
response functions strongly depend on the time elapsed since the
system was prepared in the nonequilibrium state. At equilibrium,
linear response and correlation functions are related by the
fluctuation-dissipation theorem (FDT) \cite{kubo}. Although FDT does
not hold under nonequilibrium conditions, it can be generalized by
defining an effective temperature \cite{kurchan}:

\be
T_{\rm eff}(t,t_w)=\frac{\frac{\partial C(t,t_w)}{\partial
t_w}}{R(t,t_w)} ~~~~ t\geq t_w~~~,\label{eq:teff}
\ee
where $C(t,t_w)$ is a generic two-time correlation function and
$R(t,t_w)$ is the corresponding response of the system to an
external perturbation applied at a given previous time $t_w$. At
equilibrium $T_{\rm eff}$ is just the bath temperature. But, what is
the true physical meaning of the nonequilibrium $T_{\rm
eff}(t,t_w)$? Can it be used to characterize the nonequilibrium
relaxation? Is it a well defined parameter from a thermometric point
of view? In the last years many studies have tried to answer these
questions from both empirical and theoretical perspectives. However,
there are still several debated issues (for a review see
ref.\cite{bgreviewcris} and references therein). The effective
temperature is often expressed in terms of the so-called
fluctuation-dissipation ratio (FDR):

\be
X(t,t_w)=\frac{T}{T_{\rm eff}(t,t_w)} ~~~~ t\geq
t_w~~~.\label{eq:fdr}
\ee

$X(t,t_w)=1$ for systems at equilibrium. In general, the asymptotic
value of the FDR does depend, not only on the nature of the system
but also on the type of perturbation applied~\cite{bggarrahan2}. A
property that a physically meaningful effective temperature $T_{\rm
eff}(t,t_w)$ should satisfy is its independence on the type of
observable used to define the correlation and conjugated response
functions in the limit $t\gg t_w$. A standard way to account for
possible differences is to calculate or measure $X(t,t_w)$ for
different observables to evaluate such independence.

In order to analyze the applicability and generality of the concept
of effective temperature, a variety of exactly solvable models with
glassy dynamics have been studied in the last years. A remarkable
aspect of glassy systems is the appearance of negative effective
temperatures under some conditions. This seems to contradict our
intuition and to preclude a possible thermometric interpretation of
the effective temperature. Recent studies on kinetically constrained
models reveal \textit{negative}
FDRs~\cite{sollich_negative,sollich_negative2} which have been
interpreted as due to activated effects in the dynamics of such
class of models. Negative FDRs seem to be unrelated to any
thermodynamic interpretation of effective temperatures. However,
from a theoretical point of view, nothing prevents  that they could
be generally found in glassy materials.

In the present paper we study FDRs in the context of the Backgammon
model (BG) \cite{bgfelix1}. The BG at low temperatures presents the
typical behavior of the nonequilibrium relaxation of structural
glasses: extremely slow relaxation, time-dependent hysteresis
effects, activated increase of the relaxation time and aging. The
most interesting feature of the BG is the fact that glassy behavior
is only due to the emergence of entropic barriers rather than energy
barriers.

We have found observable-independent negative FDRs in the BG due to
the entropic barriers present at low temperatures. We conclude that
the negativeness of these FDRs is a consequence of the dynamic
coupling between the external field and the energy of the system.
Interestingly, we also have found how these negative FDRs scale with
the waiting time.

The paper is organized as follows: in section~\ref{sec:model} we
briefly review the BG; in section~\ref{sec:analitico} we present the
exact analytical expressions for the correlations and responses of a
set of correlations and conjugated responses in the model; in
section~\ref{sec:results} we present both numerical and analytical
results; finally, in section~\ref{sec:conclusions} we discuss the
results. Technical aspects are left to the Appendixes.

\section{THE MODEL}
\label{sec:model}

The Backgammon Model (BG) is a mean-field model for a glass without
energy barriers. The model was introduced in \cite{bgfelix1} and has
been extensively studied in
\cite{bgfelix2,bgfelix3,bgfelix4,bggodreche1,bggodreche2,bggodreche3,bgprados}.
Similarly to the case of kinetically constrained models
\cite{bgreviewsoll}, the statics of this model is very simple and
does not show any phase transition at finite temperatures. The BG
belongs to the more general class of models called \textit{urn}
models which are based on the original Ehrenfest model
\cite{bgehrenfest,bgehrenfest2} and consist of a set of $M$ boxes
("urns") among which we can distribute $N$ particles. In these
models there is no local kinetic constraint but there exists a
conservation law, the total number of particles, that acts as a
global constraint which induces a condensation transition. For a
review of urn models and their extensions see
Refs.\cite{bookleuzzi,bgluckurn} and references therein.

Consider $N$ distinguishable particles which can occupy $M$
different boxes. Let us denote the density (number of particles per
box) by $\rho = \frac{N}{M}$. The Hamiltonian in the Backgammon
model is defined by:

\beq H=-\sum_{r=1}^M\,\de_{n_r,0} \label{bg1}~, \eeq

\noindent where $n_r$ is the occupation number of the box
$r=1,...,M$. The conservation of the number of particles gives a
global constraint:

\beq \sum_{r=1}^M\,n_r= N~~~~. \label{bg2} \eeq

Eq.(\ref{bg1}) shows that energy is simply given by the number of
empty boxes (with negative sign). The system at very low
temperatures tends to empty as many boxes as possible by
accumulating all particles in a small fraction of boxes. We define
the occupation probabilities as follows, \beq
P_k=\frac{1}{M}\,\sum_{r=1}^M\,<\de_{n_r,k}> \label{bg3}, \eeq which
is the probability of finding one box occupied by $k$ particles. In
the canonical ensemble the statics can be easily solved
\cite{bgfelix1,bgfelix2}, giving the following relation for the
occupation probabilities:

\beq P_k=\rho \frac{z^{k-1} \exp(\beta \de_{k,0})}{k!\exp(z)},
\label{eq:peq} \eeq where $z$ is the fugacity and $\beta$ is the
inverse of the temperature $T$. These quantities are related by the
condition, \beq \rho(e^{\beta}-1)=(z-\rho)e^z~~~~, \label{bg5} \eeq
expressing the fact that the density is fixed to $\rho$. This
condition, in the microcanonical formulation, is equivalent to the
saddle point condition in the integral solution of the partition
function. In the grandcanonical formulation this closure condition
is easier to obtain by means of the equation of state. The
occupation probabilities are the main observables in the system and
verify the relation $\sum_{k=0}^{\infty}P_k=1$. In particular the
energy is simply given by $U=-P_0$.

\begin{figure}[hbp!]
\begin{center}
\includegraphics*[width=14cm,height=5cm]{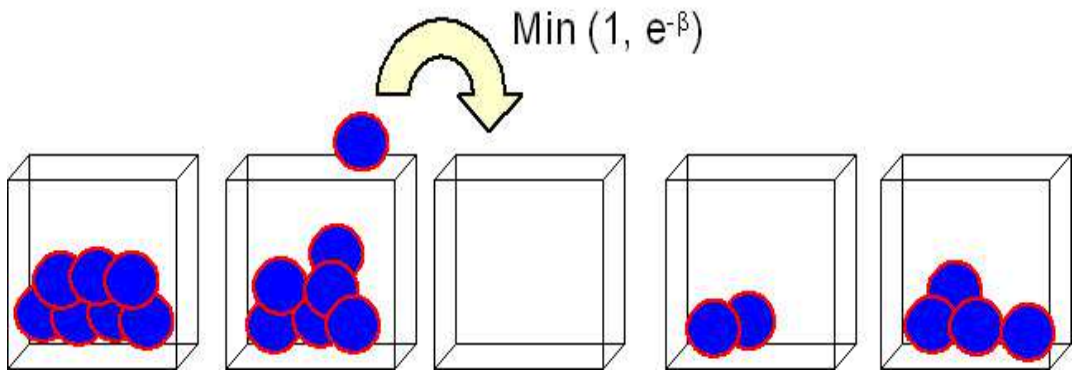}
\vskip 0.1in \caption{(Color online) Schematic representation of the
dynamics of the model. At each time step a particle is chosen at
random and a destination box selected with a uniform probability
among all boxes. In the original formulation, the system was studied
under Metropolis dynamics.} \label{fig:Bg_T}
\end{center}
\end{figure}

In the original formulation the model was studied under Metropolis
dynamics where at each time step a particle is chosen at random and
a destination box selected. The move is accepted with probability
one if the energy either decreases or does not change, and with
probability $\exp (-\beta)$ otherwise (see figure \ref{fig:Bg_T}).
Note that the energy can only increase by one unit at each time
step. The original geometry is mean-field, so the destination box is
chosen at random with uniform probability among all boxes. In this
case, a complete analytical study can be done and a hierarchy of
dynamical equations can be obtained for the occupation probabilities
\cite{bgfelix2}.

\begin{figure}[hbp!]
\begin{center}
\includegraphics*[width=14cm,height=6cm]{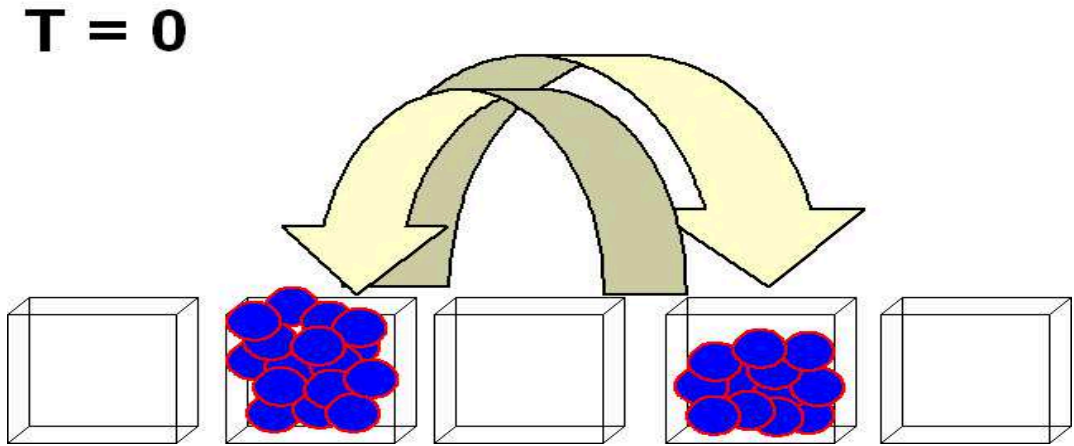}
\vskip 0.1in \caption{(Color online) At zero temperature only
movements between occupied boxes are accepted. As time goes on, only
a small fraction of boxes contain particles and the time needed to
empty an additional box increases  rapidly as the number of occupied
boxes decreases.} \label{fig:Bg_T0}
\end{center}
\end{figure}

It has been shown that the dynamics is highly non trivial at very
low temperatures where a dramatic slowing down of the relaxational
kinetics takes place. The origin of this slowing down can be
qualitatively understood. Suppose that the system starts from a
configuration of high energy and the temperature is set to zero. The
system will then evolve without accepting changes which increase the
energy of the system. As time goes on, the system evolves towards
the ground state of the system where all boxes are empty and all
particles have condensed into a single box (figure \ref{fig:Bg_T0}).
During the relaxation process more and more boxes are progressively
emptied; this means that the few boxes which contain particles have
more and more particles because the number of particles is a
conserved quantity. Then, the time needed to empty an additional box
increases with time. The final result is that the energy very slowly
converges to the ground state value. At very low temperatures it can
be shown \cite{bgfelix2} that the characteristic equilibration time
is given by: \beq \tau = t_\textrm{eq} \simeq
\frac{\exp\beta}{\beta^2} \label{eq:timebg},\eeq which diverges at
zero temperature. The Arrhenius dependence is remarkable if we note
that only entropy barriers (but not energy barriers) are present in
the model.

The BG has been used as a playground model where new concepts of
nonequilibrium thermodynamics can be tested. The fact that the
dynamics is glassy and can be exactly solved has inspired several
works that have investigated extensions of FDT to the nonequilibrium
regime (e.g. the disordered model studied in \cite{bgleuzzi}). In
the present work we solve the BG for any general Markovian dynamics
and study the existence of negative FDRs.

\bigskip

\section{CORRELATIONS AND RESPONSES IN THE BG}
\label{sec:analitico}

Let us generalize the BG by adding an external field to the
Hamiltonian of the model (\ref{bg1}). The external field is
introduced in order to compute the {\it effective temperature}, Eq.~(\ref{eq:teff}), in the nonequilibrium regime. This external field
can be a local quantity (i.e. an external field acting on a single
box) or a global one (i.e. an extensive field acting on all boxes)
leading to different definitions of the FDRs.

Previous studies of the nonequilibrium dynamics of the BG, such as
the studies carried out in~\cite{bggodreche3}, have suggested that
the effective temperature depends on the observable. In the studies
of Ref.~\cite{bggodreche3}, the Hamiltonian was perturbed by a {\it
local} external field. Recently, it has been shown that local FDRs
can lead to inconsistent results if finite-N corrections are not
properly taken into account \cite{bgferro}, pointing out the
convenience of computing {\it{global}} FDRs.

In order to give a complete picture of the system, throughout this
paper we will compute both {\it local} and {\it global} FDRs by
considering {\it local} and {\it global} external fields.

\subsection{Local external field}
\label{sec:local}

Let us consider an external field $h$ acting on a single box (e.g.
box number one) that is coupled to this box only when it contains
one particle:

\beq H=\sum_{r=1}^N\,(-\de_{n_r,0})-h\de_{n_1,1}~. \label{eq:hl}\eeq

Note that this subextensive perturbation does not affect the values
of the occupation probabilities $P_k(t)=\frac{1}{N} <\sum_r
\de_{n_r,k}>$ which in equilibrium are still given by
eq.(\ref{eq:peq}). As can be deduced from (9) we set, without loss
of generality, the density of the system $\rho = 1$. However, note
that all the results obtained throughout the
  paper are valid independently on the value of the density $\rho=N/M$
  whenever $\rho$ is finite in the $N\to\infty$ limit. In Appendix \ref{app:local} a complete
derivation of the dynamical equations for the probability densities
of the perturbed box $P_k^1$ is carried out (see
eq.(\ref{eq:appa_1})). In what follows we will focus on the
dynamical evolution of two-time quantities: local correlations and
local response functions. Local correlation functions are defined
as:

\be
C^{loc}_k(t,t_w)=\frac{1}{N}<\sum_{r}\de_{n_r(t),k}\de_{n_r(t_w),1}>~,
\label{eq:local_corr}\ee where the sum in (\ref{eq:local_corr}) runs
over all boxes and counts the fraction of boxes that contain $k$
particles at time $t$ provided that these boxes contained one
particle at previous time $t_w$. The brackets denote an average over
dynamical trajectories of the system and over the initial
conditions. The dynamical equations for these local correlations are
derived in Appendix \ref{app:local} leading to (see
eq.~\ref{eq:appa_2}):

\beqna
\frac{\partial C^{loc}_k(t,t_w)}{\partial t}&=& W(0)[-kC^{loc}_k + (k+1)C^{loc}_{k+1}-C^{loc}_{k}+C^{loc}_{k-1}]+\nn\\
&+& (W(0)-W(-1))[P_1(C^{loc}_k-C^{loc}_{k-1})+(\de_{k,1}-\de_{k,0})(C^{loc}_1(1-P_0)+C^{loc}_0P_1)]+\nn\\
&+&
(W(0)-W(1))[P_0(kC^{loc}_k-(k+1)C^{loc}_{k+1})+(\de_{k,0}-\de_{k,1})(C^{loc}_0(1-P_1)+C^{loc}_1P_0)]~.\nn\\
\label{eq:corr_loc}\eeqna

This expression is valid for any Markovian dynamics. $W(\Delta E)$
denotes the transition probability between two states with energy
difference $\Delta E$. From now on, we consider that the dynamics
obeys local detailed balance in order to ensure the convergence
toward equilibrium.

Similarly, we can compute the dynamical equations for the local
response function defined as the variation of the occupation
probabilities for the perturbed box when the impulse field is
applied at $t_w$:

\be R^{loc}_k(t,t_w)=\left( \frac{\delta P_k^1(t)}{\delta h(t_w)}
\right)_{h(t_w)\rightarrow 0}~.\ee

Again, the details about the derivation can be found in the Appendix
\ref{app:local}. The result (eq.(\ref{eq:appa_3})) is:

\beqna \frac{\partial R^{loc}_k(t,t_w)}{\partial t}&=&
W(0)[-kR^{loc}_k +
(k+1)R^{loc}_{k+1}-R^{loc}_{k}+R^{loc}_{k-1}]\nn\\
&+&(W(0)-W(-1))[P_1(R^{loc}_k-R^{loc}_{k-1})+(\de_{k,1}-\de_{k,0})(R^{loc}_1(1-P_0)+R^{loc}_0P_1)]\nn\\
&+&(W(0)-W(1))[P_0(kR^{loc}_k-(k+1)R^{loc}_{k+1})+(\de_{k,0}-\de_{k,1})(R^{loc}_0(1-P_1)+R^{loc}_1P_0)]\nn\\
&+&\de(t-t_w)S^{loc}[<P_k>]~, \label{eq:resp_loc} \eeqna

\noindent where the $\delta$-term $S^{loc}[<P_k>]$ is given in
eq.(\ref{eq:appa_delta}). Equations (\ref{eq:corr_loc}) and
(\ref{eq:resp_loc}) are the necessary ingredients for computing
nonequilibrium effective temperatures.

From (\ref{eq:corr_loc}) and (\ref{eq:resp_loc}), we can check that
FDT is verified at equilibrium. Indeed, at equilibrium the
correlations and responses become functions of the difference of
times, i.e $C^{loc}_k(t-t_w)$ and $R^{loc}_k(t-t_w)$ so we recover
time-translational invariance. Moreover, as we can see from the form
of the dynamical equations for the autocorrelations
(eq.(\ref{eq:corr_loc})) and responses (eq.(\ref{eq:resp_loc})), at
equilibrium the FDT is verified at all times provided that the
initial condition for the responses (the function $S^{loc}[<P_k>]$)
corresponds to the value of the derivative of the appropriate
correlation at equal times. In this case, the correlation functions
for a general observable are given by: \be
C^{loc}_k(t_w,t_w)=P_1(t_w)\de_{k,1}~. \ee Therefore, the initial
value ($t=t_w$) for the derivative of the correlation functions are:

\beqna \left( \frac{\partial C^{loc}_k(t,t_w)}{\partial t}
\right)_{t\rightarrow t_w}
&=& P_1(t_w)\left[ W(0)(-2\de_{k,1}+\de_{k,0}+\de_{k,2})\right]+\nn\\
&+&P_1(t_w)\left[W(0)-W(-1)\right](P_1(\de_{k,1}+\de_{k,2}))+\nn\\
&+&P_1(t_w)\left[W(0)-W(-1)\right](1-P_0)(\de_{k,1}+\de_{k,0})~.
\eeqna

Using eq.(\ref{eq:appa_delta}) it is easy to check that, in
equilibrium, FDT is verified. \be T=-\frac{\frac{\partial
C^{loc}_k(t-t_w)}{\partial t}}{R^{loc}_k(t-t_w)}~. \ee

\subsection{Global external field}
\label{sec:global}

As we have explained before, local computations can lead to
erroneous conclusions if finite-$N$ corrections are not properly
taken into account \cite{bgferro}. In such cases it is easier to
carry out an analysis of FDRs for global observables. Here we shall
consider the corresponding extensive perturbation of an external
field coupled to the set of boxes which contain just one particle
(i.e coupled to the observable $P_1$). The Hamiltonian reads:

\be H=-\sum_{r=1}^N\,(\de_{n_r,0}+h\de_{n_r,1}) ~.\ee

Now, as the perturbation is extensive, all the equilibrium
occupation probabilities are modified in the presence of the
external field $h$:

\be P_k=\frac{z^{k-1} \exp(\beta\de_{k,0}-\beta h \de_{k,1})}{k!
(e^z+e^{-\beta h}-1)}~. \label{eq:peqh}\ee

We proceed following the same steps as in the local case. In
Appendix \ref{app:global} we have computed the dynamical equations
for the occupation probabilities, eq.(\ref{eq:appb_1}), and from
these equations we derive the dynamical evolution for the global
correlation and response functions.

Due to the fact that the perturbation is extensive we consider the
connected correlation functions:

\be C^g_k(t,t_w)=<\gamma_k(t)\gamma_1(t_w)>~, \ee where
 \be
\gamma_k(t)=\frac{1}{N} \sum_r \de_{n_r,k} - P_k(t)~ \ee are the
deviations of the instantaneous occupation variables from their
average value at a given time.  The dynamical evolution for the
global correlation functions is given by eq.(\ref{eq:appb_2}):

\beqna
\frac{\partial C^g_k(t,t_w)}{\partial t} &=& W(0)[-kC^g_k + (k+1)C^g_{k+1}-C^g_{k}+C^g_{k-1}]+\nn\\
&+& (W(0)-W(-1))[C^g_1(\de_{k,1}-\de_{k,0}+P_k-P_{k-1})+P_1(C^g_k-C^g_{k-1})]+\nn\\
&+&
(W(0)-W(1))[C^g_0(kP_k-(k+1)P_{k+1}+\de_{k,0}-\de_{k,1})+P_0(kC^g_k-(k+1)C^g_{k+1})].\nn\\
\label{eq:global_corr}\eeqna Again, these equations are valid for
any Markovian dynamics. The global response function is the response
of the occupation probabilities to the extensive perturbation
coupled to the observable $P_1$:

\be R^g_k(t,t_w)=\left( \frac{\delta P_k(t)}{\delta h(t_w)}
\right)_{h(t_w)\rightarrow 0}~. \ee

The result for the dynamical evolution is given in Appendix
\ref{app:global}, eq.(\ref{eq:appb_3}), and it reads:

\beqna \frac{\partial R^g_k(t,t_w)}{\partial t} &=& W(0)[-kR^g_k +
(k+1)R^g_{k+1}-R^g_{k}+R^g_{k-1}]+\nn\\ &+&
(W(0)-W(-1))[R^g_1(\de_{k,1}-\de_{k,0}+P_k-P_{k-1})+P_1(R^g_k-R^g_{k-1})]+\nn\\
&+&
(W(0)-W(1))[R^g_0(kP_k-(k+1)P_{k+1}+\de_{k,0}-\de_{k,1})+P_0(kR^g_k-(k+1)R^g_{k+1})]+\nn\\
&+& \de(t-t_w) S^g[<P_k>]~,\label{eq:global_resp}\eeqna where we
have introduced the function $S^g[<P_k>]$ which depends only on one
time and gives the initial value for the responses. The exact form
of $S^g[<P_k>]$ is given in eq.~(\ref{eq:appb_delta}).

Again, we check that in equilibrium FDT is verified. Indeed, at
equal times the global correlations are given by:

\be C^g_k(t_w,t_w) = - P_k(t_w)(P_1(t_w)-\de_{k,1})~.
\label{eq:correqtimes}\ee

Obviously, in equilibrium the correlations at equal times do not
depend on time. Inserting this initial value in the equations for
the correlations and by considering the equilibrium case, it is easy
to prove FDT for all $k$ values of the observables $C^g_k$, $R^g_k$:

\be T=-\frac{\frac{\partial C^g_k(t-t_w)}{\partial t}}{R^g_k(t-t_w)}
~. \ee

\section{RESULTS}
\label{sec:results}

In this section we analyze the nonequilibrium behavior of the
correlations and responses at zero temperature both for local and
global observables.

The interesting glassy behavior in the BG model occurs in the zero
temperature limit where entropy barriers govern the relaxational
dynamics of the model. In what follows we shall consider heat-bath
dynamics at zero temperature, both for the local and the global
variables. This choice is motivated by the known fact that in the
Metropolis algorithm there is a discontinuity  of the derivative of
the transition rates for $\Delta E=0$. As a result, the definition
of the response functions become ambiguous, see
Ref.\cite{bggodreche3}. We circumvent this drawback by employing
heat bath dynamics.

\subsection{Local two-time quantities}

From the numerical integration of equations (\ref{eq:corr_loc}) and
(\ref{eq:resp_loc}) we can analyze the nonequilibrium behavior of
the local correlations and response functions. From now on, all the
numerical results shown are obtained using heat bath dynamics at
zero temperature.

\subsubsection{Correlations and responses}

\begin{figure}[hbp!]
\begin{center}
\includegraphics*[width=10cm,height=8cm]{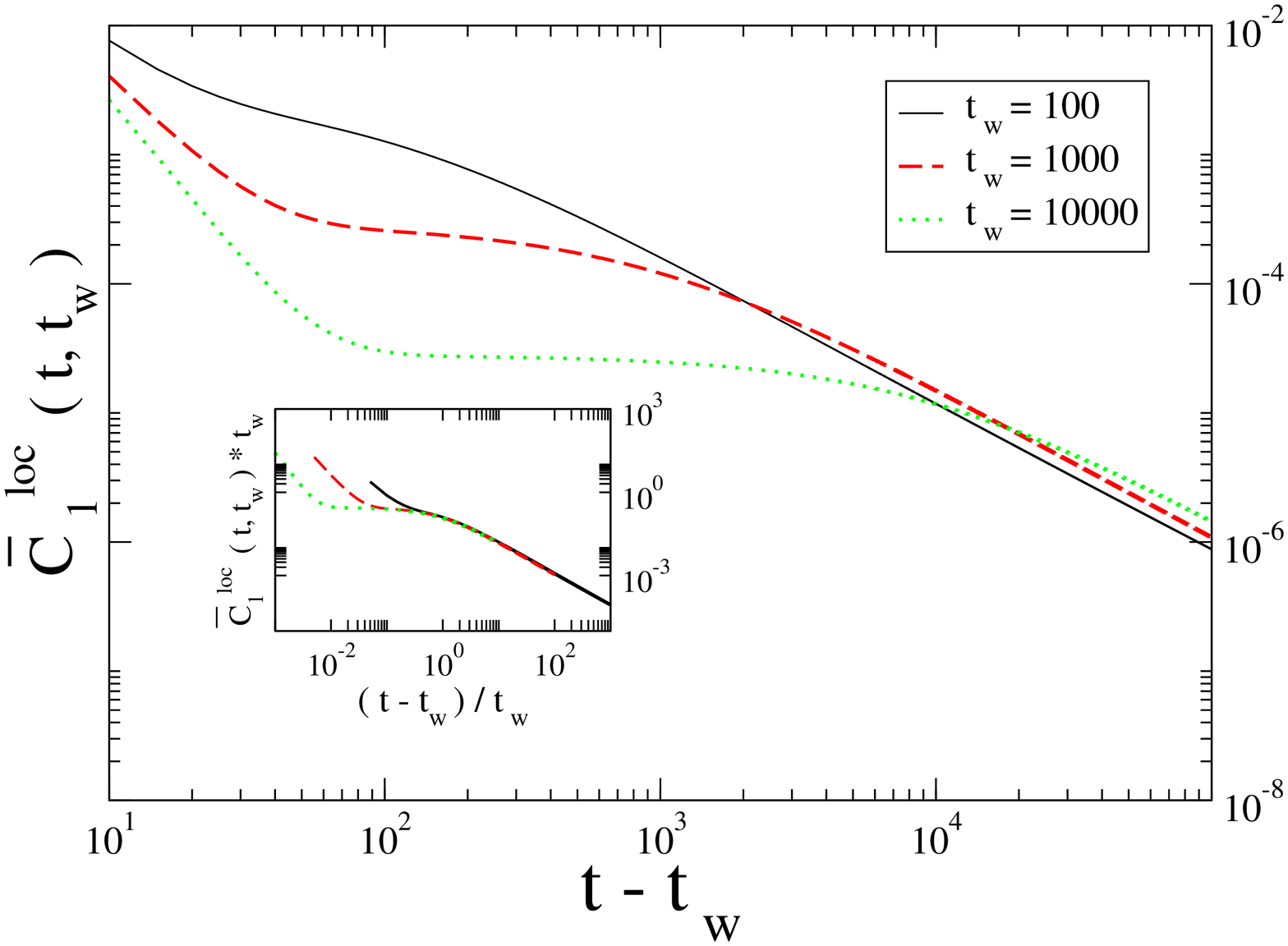}
\vskip 0.1in \caption{(Color online) The evolution for the
normalized local correlation function $\bar{C}^{loc}_1(t,t_w)$
(dimensionless) for different $t_w$'s. In the inset the scaling of
$\bar{C}^{loc}_1(t,t_w)$ corresponding to simple aging is displayed.
The time is measured in Monte Carlo sweeps.} \label{fig:corr_loc}
\end{center}
\end{figure}

In figure \ref{fig:corr_loc} we plot the normalized local
correlation $\bar{C}^{loc}_1(t,t_w)=\frac{C_1(t,t_w)}{P_1(t_w)}$ at
zero temperature. We can clearly see the aging effects in the local
correlation function: as $t_w$ increases the autocorrelation
function develops a plateau showing two characteristic and well
separated time-scales: the first timescale corresponds to the
initial relaxation of the system (usually called $\beta$-relaxation)
which does not depend much on $t_w$; the second one is larger,
increases with $t_w$ and corresponds to the late decay of the
correlation function, usually known as $\alpha$-relaxation. The
existence of these two time-scales is a typical signature of the
glassy relaxation of structural glasses.

In the inset of figure \ref{fig:corr_loc} we plot the local
normalized correlation function $\bar{C}^{loc}_1(t,t_w)$ multiplied
by $t_w$ in order to collapse all curves on the same plateau. It is
clear that the system displays simple aging, i.e the scaling $t/t_w$
is well satisfied.

\begin{figure}[hbp!]
\begin{center}
\includegraphics*[width=10cm,height=8cm]{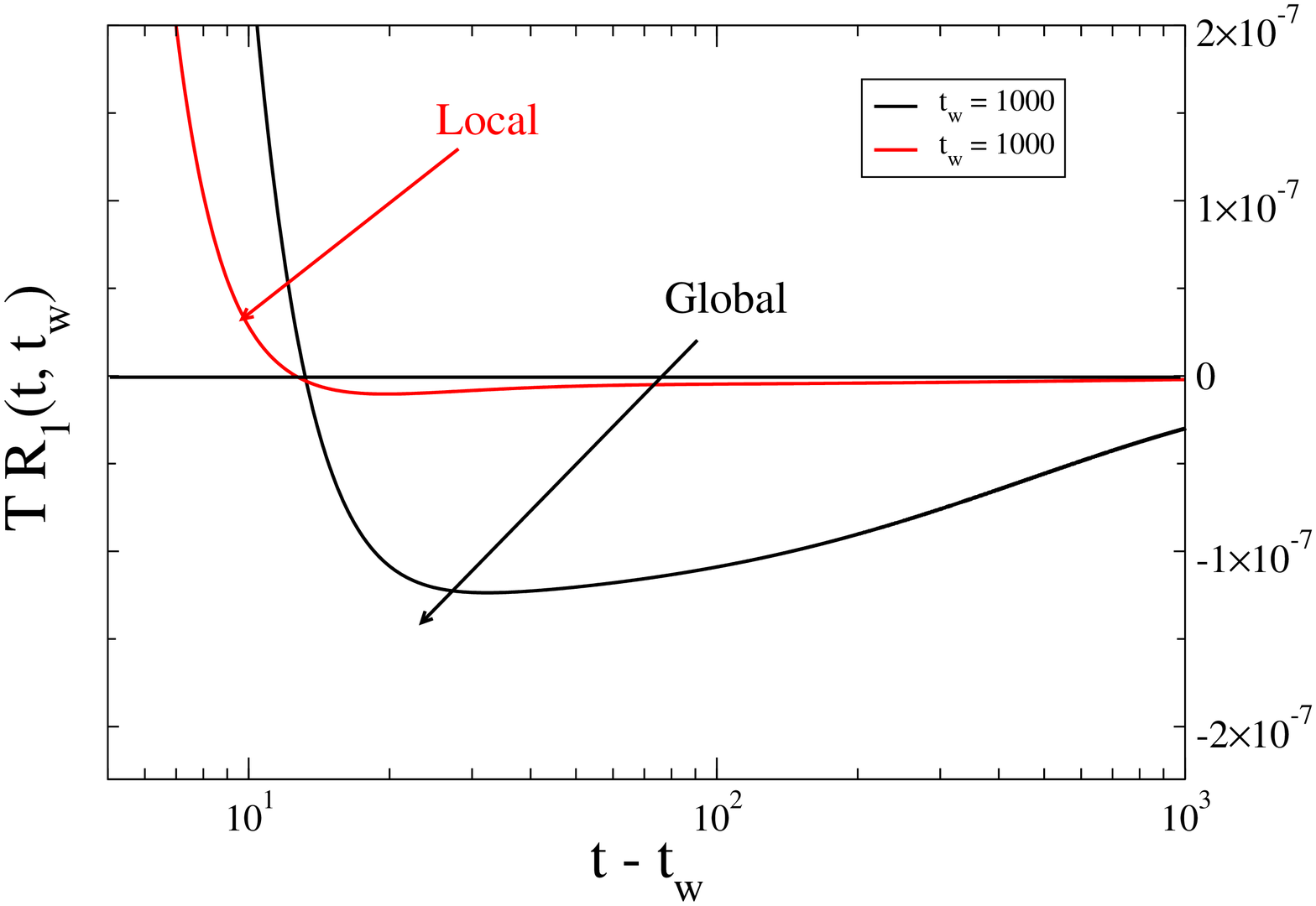}
\vskip 0.1in \caption{(Color online) Time dependence of the global
and local dimensionless response functions multiplied by $T$
($TR^{g}_1$ and $TR^{loc}_1$) at $t_w = 1000$. The time is measured
in Monte Carlo sweeps.} \label{fig:resp1}
\end{center}
\end{figure}

Regarding response functions, they show some peculiarities: on the
one hand, the initial value for the response functions (given by the
function $S^{loc}[<P_k>]$) is proportional to $\beta$, giving a
divergence at zero temperature (a known common feature of
kinetically constrained models \cite{sollich_negative2}); on the
other hand, the response function $R^{loc}_1(t,t_w)$ is
non-monotonic (for $t_w$ fixed  when $t$ is varied) and becomes
negative for long enough times. In figure \ref{fig:resp1} we plot
both the local, $R^{loc}_1$, and the global, $R^g_1$, response
functions for the observable $P_1$ (see below). Both responses show
a non-monotonic behavior and become negative for long times.

The non-monotonicity of the response function can be easily
understood. The external field is coupled to $P_1$, therefore the
system tends to increase the population of boxes with one particle;
however, because boxes with one particle are bottlenecks for the
relaxation of the energy, a transient increase in their number at
$t_w$, induces a faster relaxation of the energy at later times.
Because the natural evolution of the system tends to decrease $P_1$
when decreasing the energy, a transient increase of $P_1$ at $t_w$
causes a net decrease of the same quantity at later times when
energy relaxation becomes faster.

In order to facilitate the readings of the effective temperature in
our plots we introduce the function $\bar{G}^{loc}_1(t,t_w)$ as:
\be\bar{G}^{loc}_1(t,t_w)=T\frac{|(R^{loc}_1(t,t_w))|}{P_1(t_w)}~~,\label{eq:gloc}\ee
which is the normalized absolute value of the response
$R^{loc}_1(t,t_w)$ multiplied by $T$. Due to the change of sign of
$R^{loc}_1$, we have taken the absolute value in order to plot the
relaxation in a log-log scale.

\begin{figure}[tbp]
\begin{center}
\includegraphics*[width=10cm,height=8cm]{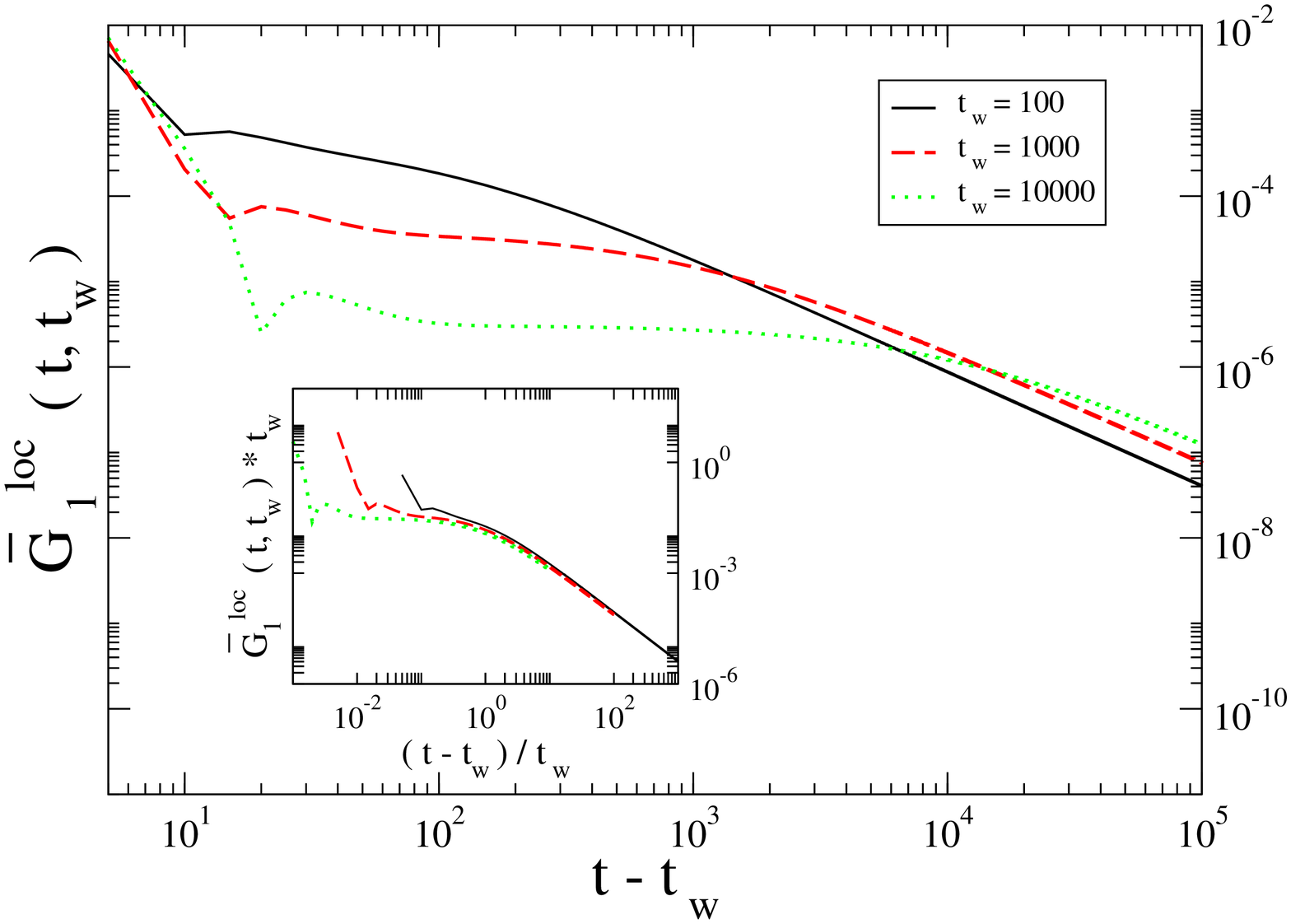}
\vskip 0.1in \caption{(Color online) The evolution of the
dimensionless quantity, $\bar{G}^{loc}_1(t,t_w)$, (\ref{eq:gloc}) at
different values of $t_w$. In the inset we show the simple aging
scaling for this quantity. The dips observed in $\bar{G}^{loc}_1$
indicate changes of sign in $R^{loc}_1$. The time is measured in
Monte Carlo sweeps.\label{fig:resp_loc}}
\end{center}
\end{figure}

In figure \ref{fig:resp_loc} we plot $\bar{G}^{loc}_1(t,t_w)$ for
different values of $t_w$. As can be inferred from figure
\ref{fig:resp_loc}, the dip at short times corresponds to the change
in sign of the response. Looking at this logarithmic plot, the
response shows again the two characteristic relaxation time-scales
of glasses. In the inset of figure \ref{fig:resp_loc} we can see
that the response function also displays simple aging with scaling
$t/t_w$ as the leading term.

It is worth noting that this simple aging relaxation in the
$\alpha$-regime can also be seen in the correlations and responses
for the other observable quantities of the model, i.e in the
dynamical behavior of $C^{loc}_k(t,t_w)$ and $R^{loc}_k(t,t_w)$ for
a generic $k$ (data not shown).

\subsubsection{Nonequilibrium effective temperatures}

\bigskip
\bigskip

We now define a set of effective temperatures from the
nonequilibrium definition, eq.(\ref{eq:teff}):

\be \left(T_{\rm eff}^{loc}\right)_k(t,t_w)=\frac{ \frac{\partial
C^{loc}_k(t,t_w)}{\partial t_w}}{R^{loc}_k(t,t_w)}~. \ee

\begin{figure}[tbp]
\begin{center}
\includegraphics*[width=10cm,height=8cm]{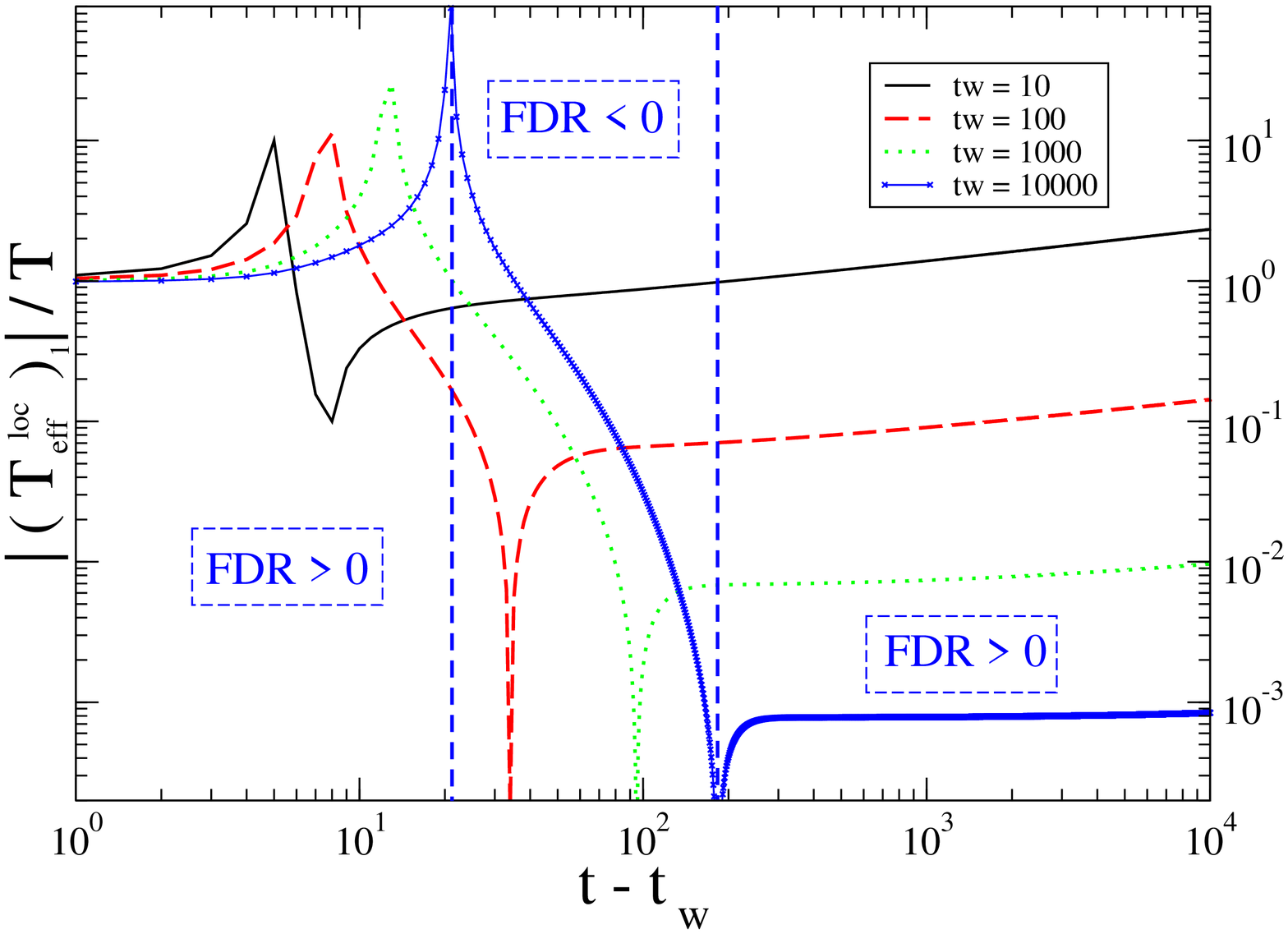}
\vskip 0.1in \caption{(Color online) The absolute value of the
effective temperature divided by $T$ for the observable $P_1$
($k=1$) as a function of time for different $t_w$'s. The (up and
down) oriented
  spikes indicate changes in the sign of $\left(T_{\rm
    eff}^{loc}\right)_1$. Note that, for a given $t_w$, the effective
temperature changes sign twice. Therefore, we can distinguish three
regions depending of the sign of the local FDR. The time is measured
in Monte Carlo sweeps.\label{fig:teff_loc}}
\end{center}
\end{figure}

In order for $\left(T_{\rm eff}^{loc}\right)_k(t,t_w)$ to share some
of the properties of a thermometric temperature it should not
asymptotically depend on the integer $k$ (for a fixed $t_w$ and in
the large $t$ limit). In figure \ref{fig:teff_loc} we plot the ratio
between the absolute value of the effective temperature and the
physical one (in the limit $T\rightarrow 0$), which corresponds to
the inverse of the local FDR~\cite{kurchan} defined as:

\be X^{loc}_k(t,t_w)=\frac{T}{\left(T_{\rm
eff}^{loc}\right)_k(t,t_w)}~~~~.\ee

We can clearly see that the effective temperature shows two
different behaviors depending on the time-scales considered. For
$t\to t_w$ the value of the FDR converges to 1 as $t_w$ increases.
This is a typical feature of glasses: the first $\beta$-relaxation
is an equilibrium process which implies that the effective
temperature is just the physical one. This is true in the asymptotic
limit $t_w\rightarrow\infty$. It can be shown that it converges to 1
in a logarithmic way as was found in the analysis of
Ref.\cite{bggodreche3}.

From the integration of the dynamical equations, we obtain the
asymptotic value of the ratio $\left(T^{loc}_{\rm
eff}\right)_k(t,t_w)/T$, which is positive because for long enough
times both the local response and the derivative of the local
correlation become negative. This asymptotic value tends to zero in
the limit $t_w \rightharpoonup\infty$. In addition, for a given
waiting time the effective temperature is proportional to the bath
temperature.

\begin{figure}[tbp]
\begin{center}
\includegraphics*[width=10cm,height=8cm]{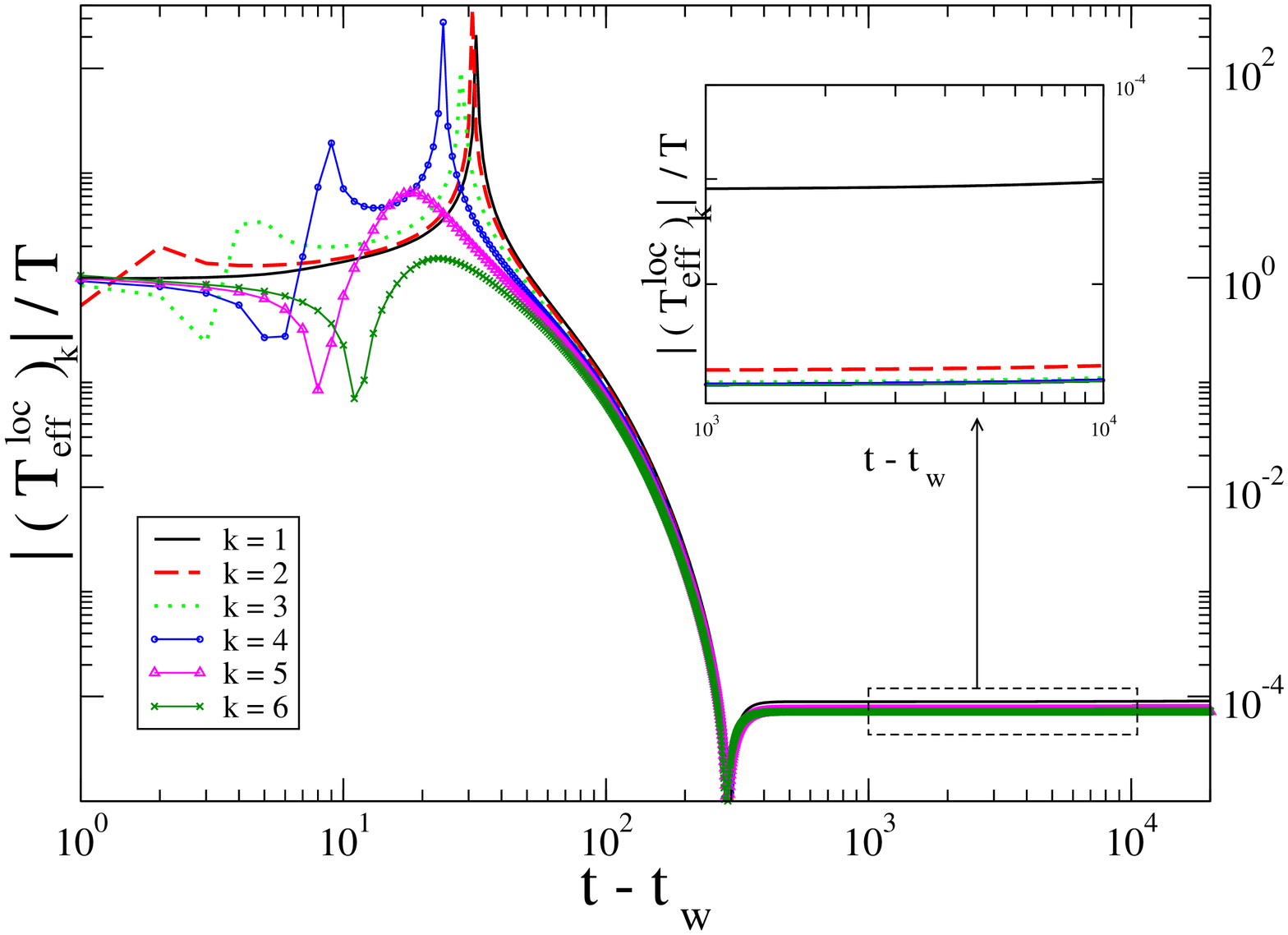}
\vskip 0.1in \caption{(Color online) The absolute value of the
effective temperature divided by $T$ ($t_w=10000$) as a function of
time for different observables $k$. In the inset we zoom the boxed
part of the figure in order to emphasize the observable dependence
of the effective temperature. The (up and down) oriented spikes
indicate changes in the sign of $\left(T_{\rm eff}^{loc}\right)_k$.
The time is measured in Monte Carlo
sweeps.}\label{fig:teff_comp_loc}
\end{center}
\end{figure}

Looking at figure \ref{fig:teff_comp_loc}, where we plot the
effective temperature at $t_w=10000$ for different observables, we
can clearly see that the effective temperature depends on the
observable under scrutinity. Consequently, it seems clear that from
a local point of view we cannot define a unique effective
temperature by using the FDR.

\subsection{Global two-time quantities}

In the preceding section we have shown that a unique effective
temperature cannot be defined by the FDR from a local perturbation.
As we have mentioned before, this is an expected result consistent
with previous analysis~\cite{bggodreche3}. In this section we will
analyze the time dependence of the global correlation and the global
response functions and we will show that a {\it{unique}} negative
effective temperature can be defined from the global FDRs.

\bigskip

\subsubsection{Correlations and responses}

\bigskip

\begin{figure}[tbp]
\begin{center}
\includegraphics*[width=10cm,height=8cm]{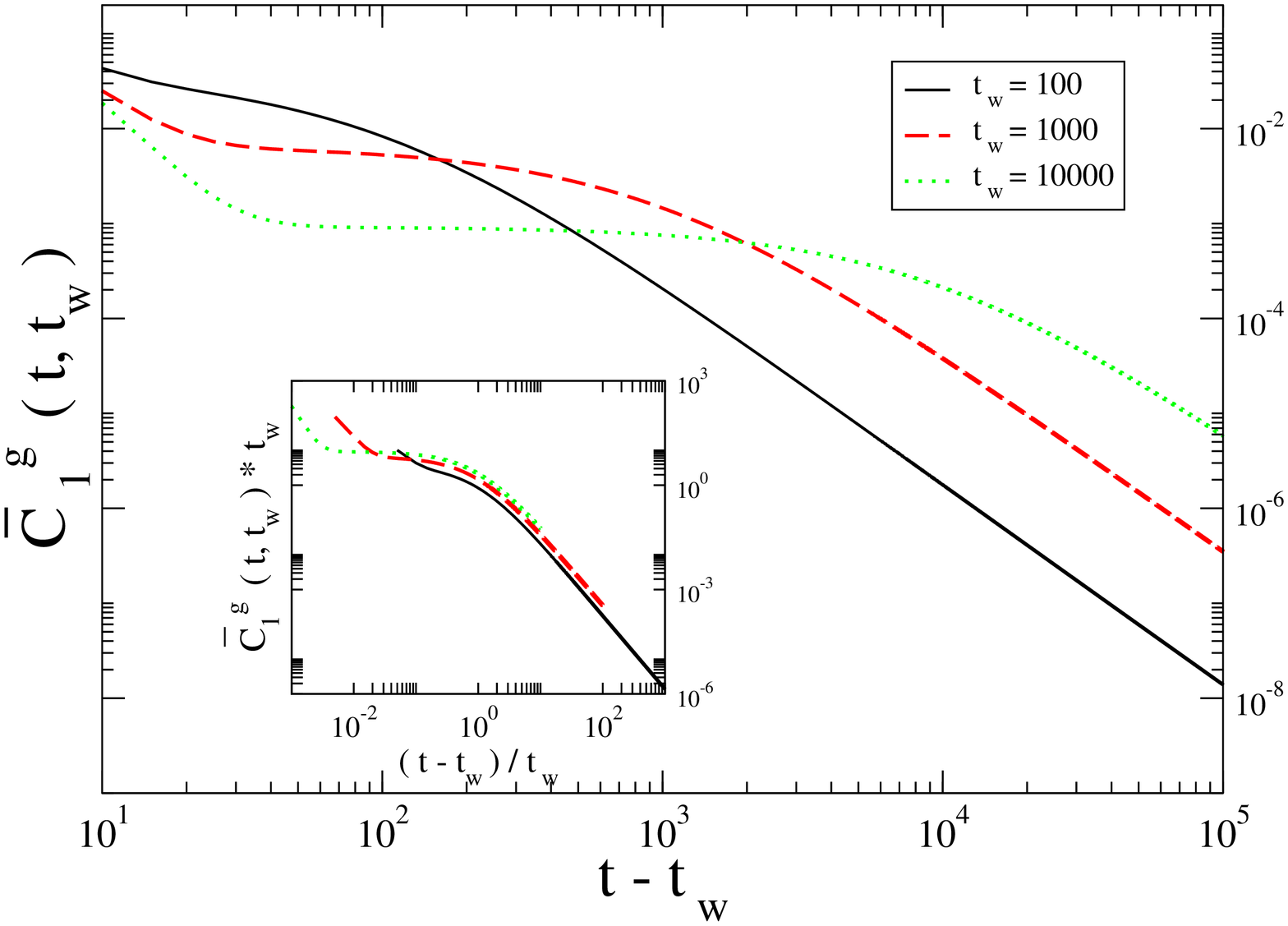}
\vskip 0.1in \caption{(Color online) The global (connected)
correlation function normalized for different
$t_w$'s\label{fig:corr_glob}. In the inset we plot the simple aging
scaling for $\bar{C}^g_1(t,t_w)$. These correlation functions are
dimensionless and the time is measured in Monte Carlo sweeps.}
\end{center}
\end{figure}

We study the connected correlation functions for heat bath dynamics
of the BG at zero temperature. In Fig.\ref{fig:corr_glob} we plot
the normalized correlation function,
$\bar{C}^g_1(t,t_w)=\frac{C^g_1(t,t_w)}{P_1(t_w)}$, for different
values of $t_w$. Similarly to the local case, we can clearly
distinguish two characteristic time-scales in the system, the
$\beta$-relaxation and the $\alpha$-relaxation. Note that, as $t_w$
increases, the plateau value of the correlation decreases and in the
limit $t_w\to\infty$ the plateau value converges to zero. In the
inset of Fig.\ref{fig:corr_glob} we have multiplied this normalized
correlation by $t_w$. As for the local case, the global correlation
displays simple aging.

\begin{figure}[tbp]
\begin{center}
\includegraphics*[width=10cm,height=8cm]{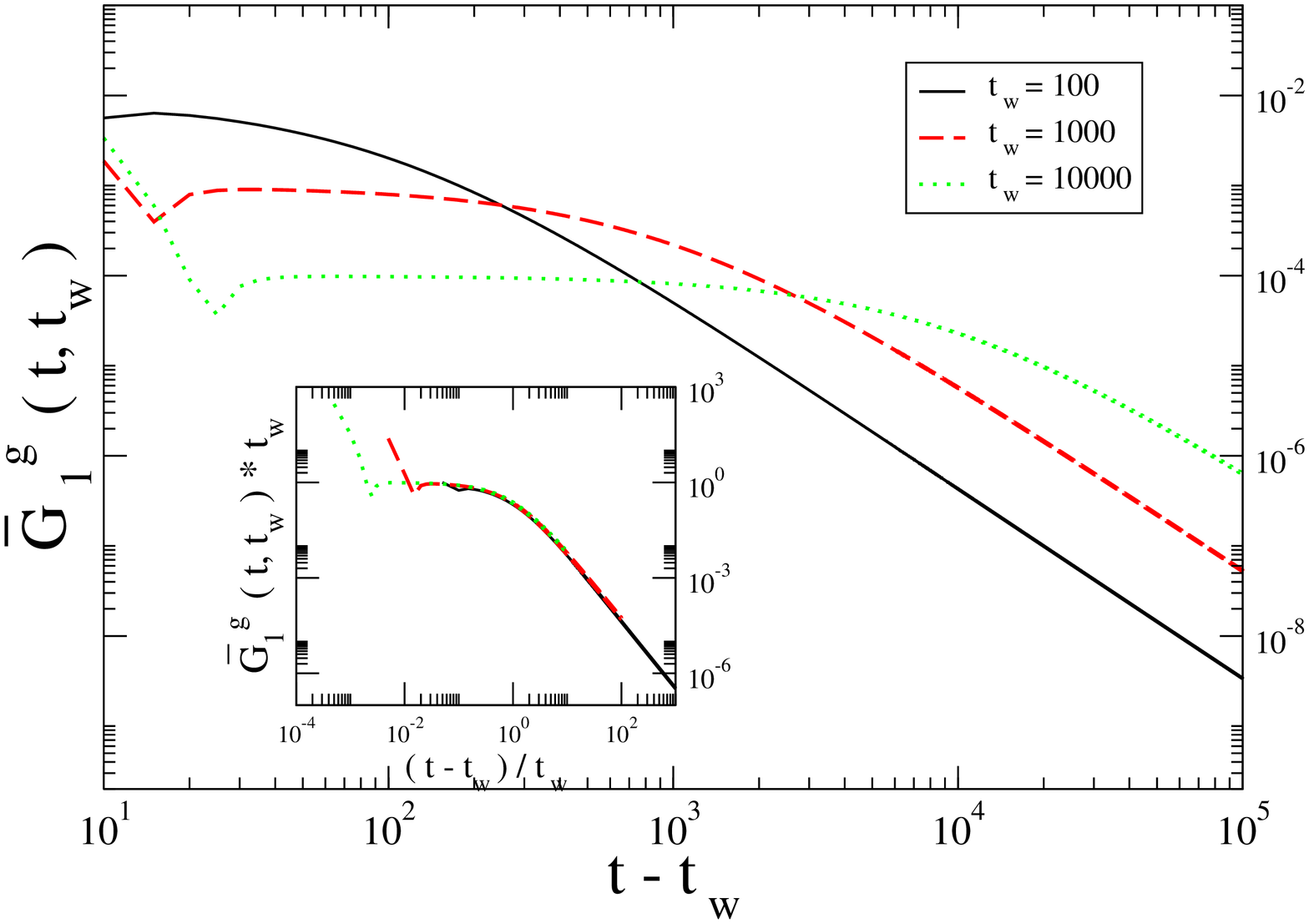}
\vskip 0.1in \caption{(Color online) The global scaled response
function for $P_1$ for different $t_w$'s. The dips observed in
$\bar{G}^{g}_1$ (dimensionless) indicate changes in the sign of
$R^{g}_1$. The time is measured in Monte Carlo
sweeps.}\label{fig:resp_glob}
\end{center}
\end{figure}

Again, in order to analyze the relaxation of the global response
function we have defined the normalized response function
$\bar{G}^{g}_1(t,t_w)$

\be \bar{G}^{g}_1(t,t_w)=T\frac{|(R^{g}_1(t,t_w))|}{P_1(t_w)}, \ee
motivated by the fact that the response function $R_1^g(t,t_w)$
becomes negative for long times as it is shown in figure
\ref{fig:resp1}. This is again consequence of the fact that the
natural evolution of the system tends to diminish $P_1(t_w)$ in
opposition to the action of the external field. Moreover, the global
response function is proportional to the bath temperature, which
diverges at zero temperature. In figure \ref{fig:resp_glob} we plot
the two time-scales relaxation of $\bar{G}^{g}_1(t,t_w)$. In the
inset of figure \ref{fig:resp_glob} we show the simple aging scaling
of the function $\bar{G}^{g}_1(t,t_w)$. Again, the dip of the curves
at short times corresponds to the time when the response changes its
sign.

The global correlations and responses for the rest of the
observables in the model, $\bar{C}_k^g(t,t_w)$ and
$\bar{G}_k^g(t,t_w)$, also display simple aging (data not shown). It
is worth mentioning that the ratio between $\bar{C}_k^g$ and
$\bar{G}_k^g$ for $k>1$ is of the same order of magnitude as the one
corresponding to $k=1$.

\subsubsection{Nonequilibrium effective temperatures}

\bigskip

As we have done for local observables, from the FDR we can define
the effective temperatures:

\be \left(T_{\rm eff}^{g}\right)_k(t,t_w)=\frac{ \frac{\partial
C^{g}_k(t,t_w)}{\partial t_w}}{R^{g}_k(t,t_w)}~. \ee

\begin{figure}[tbp]
\begin{center}
\includegraphics*[width=10cm,height=8cm]{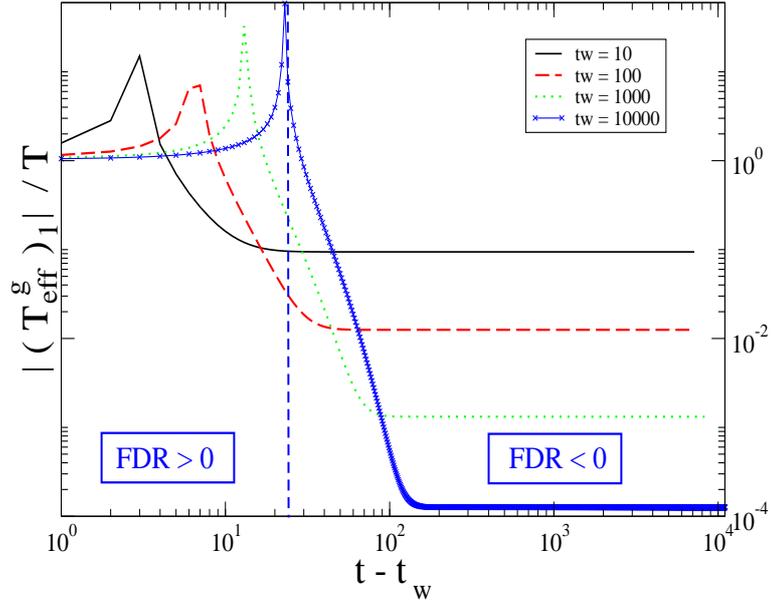}
\vskip 0.1in \caption{(Color online) The absolute value of the
global effective temperature function normalized by $T$ for $P_1$
for different values of $t_w$. The up oriented spikes indicate
changes in the sign of $\left(T_{\rm
    eff}^{g}\right)_1$ which is negative for long times. The time is measured in Monte Carlo sweeps.\label{fig:teff_glob}}
\end{center}
\end{figure}

In figure \ref{fig:teff_glob} we plot the absolute value of the
effective temperature $\left(T^{g}_{\rm eff}\right)_1(t,t_w)$
divided by $T$ for different values of $t_w$. This quantity is
related to the global FDR $X^g_1(t_w)$ for $P_1$ as:

\be X^g_1(t_w)= \frac{T}{\left(T^{g}_{\rm
eff}\right)_1(t,t_w)}~~,\label{eq:globfdr}\ee

As in the local case, the value of $\frac{\left(T^{g}_{\rm
eff}\right)_1(t,t_w)}{T}$ in the limit $t\to t_w$ tends to 1 showing
that the first $\beta$-regime corresponds to an equilibrium
relaxation process (i.e $X^g_1=1$). We can also see that, in
contrast with the local case,  this global effective temperature
remains constant throughout the $\alpha$-regime for any finite
$t_w$.

\begin{figure}[tbp]
\begin{center}
\includegraphics*[width=10cm,height=8cm]{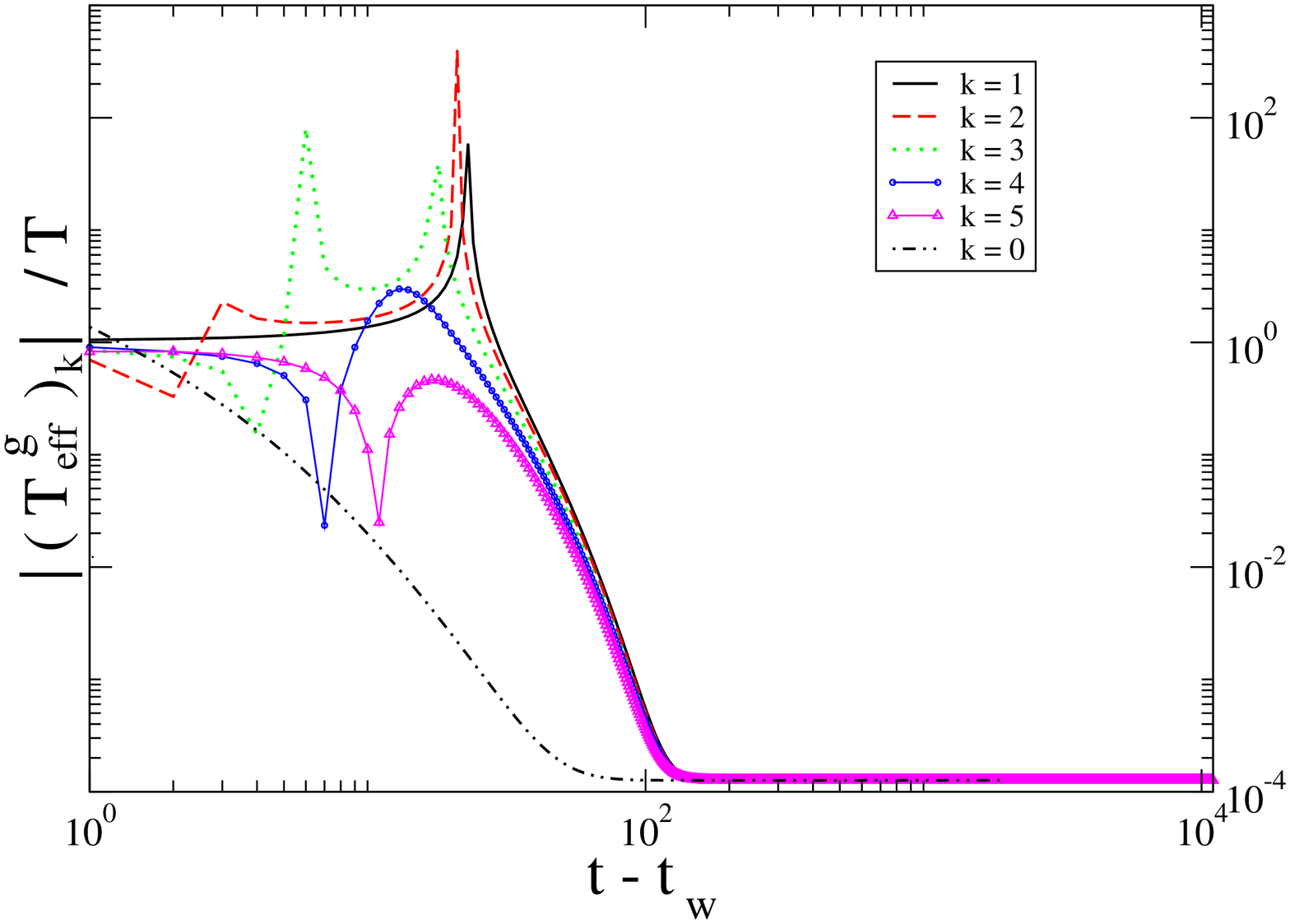}
\vskip 0.1in \caption{(Color online) The global effective
temperature function normalized by $T$ for $tw=10000$ and for
different observables. The (up and down) oriented
  spikes indicate changes in the sign of   $\left(T_{\rm
    eff}^{g}\right)_k$. Note that for some values of $k$ the effective
temperature changes sign more than twice. The time is measured in
Monte Carlo sweeps.\label{fig:teff_comp_glob}}
\end{center}
\end{figure}

A very important aspect of the global effective temperature is the
fact that, for a given value of the waiting time this effective
temperature does not depend on the observable as can be seen in
figure \ref{fig:teff_comp_glob} where we have plotted the absolute
value of the inverse of the global FDRs (eq.(\ref{eq:globfdr})) at
$t_w=10000$ for different observables. It is clear that the
asymptotic value of the FDRs at finite $t_w$ does not depend on the
observable.

Moreover, from the results of figure \ref{fig:teff_glob} we can see
that for large waiting times $t_w$ the inverse of the FDR scales as
the inverse of $t_w$:

\be \frac{\left(T^{g}_{\rm
eff}\right)_k(t,t_w)}{T}=\frac{1}{X^g_k(t_w)}\simeq -\frac{1}{t_w}
~~~~ \forall ~k~.\label{eq:teff_sca}\ee

Again, the minus sign in eq.(\ref{eq:teff_sca}) is a consequence of
the non-monotonicity of the response functions. Finally, it is worth
mentioning that we have checked that all the results obtained at
zero temperature throughout this paper remain valid at finite but
very low temperatures. The analysis at finite small temperatures
does not give new insights into the nonequilibrium behavior of the
system as all dynamical quantities smoothly converge to their $T=0$
limit.

\subsection{Asymptotic analysis}

In the preceding section we have obtained a negative FDR independent
of the observable that displays simple scaling of the type $t/t_w$.
This result can be easily understood by analyzing the asymptotic
nonequilibrium relaxation of the model. The equilibrium
probabilities in the presence of an external field are given by:
\beq P_k=\frac{z^{k-1} \exp(\beta
(\de_{k,0}-h\de_{k,1})}{k!\exp(z)}.\eeq In the long time asymptotic
regime, the multiplier $z(t)$ is a function of time which grows
as~\cite{bgfelix3,bgfelix4}:

\beq z(t)\approx \ln t + \ln(\ln t)~~.\label{eq:zas}\eeq

With the global perturbation considered along the paper we can
compute the global susceptibility $\chi^g_1$ by assuming local
equilibrium using (\ref{eq:peqh}) with $k=1$:

\beq T\chi^g_1=T\lim_{h\rightarrow 0}
\frac{P_1(h=0)-P_1(h)}{h}=\frac{1}{e^z}~~.\eeq

From (\ref{eq:zas}) the global susceptibility associated to the
observable $P_1(t)$ decays as:

\beq T\chi^g_1(\tau)=\frac{1}{\tau \ln \tau}~~,\eeq where
$\tau=t-t_w$. The asymptotic decay of $R^g_1(\tau)$ is given by the
derivative of $\chi^g_1(\tau)$ multiplied by the temperature:

\beq TR^g_1(\tau) =  -\frac{1}{\tau^2\ln \tau}+
{\cal{O}}\left(\frac{1}{\tau^2\ln^2\tau}\right)
~~.\label{eq:r1_sca}\eeq

Now, by using the equation (\ref{eq:global_resp}) at zero
temperature:

\be \frac{\partial R_0^g}{\partial \tau}=
R^g_1-P_0R_1^g-P_1R_0^g~~~~,\ee

we obtain the asymptotic decay of $R^g_0$:

\be TR^g_0(\tau) = \frac{1}{\tau\ln^2\tau}+
{\cal{O}}\left(\frac{1}{\tau^2\ln^3\tau}\right) ~~.
\label{eq:r0_sca}\ee

\begin{figure}[tbp]
\begin{center}
\includegraphics*[width=14cm,height=10cm]{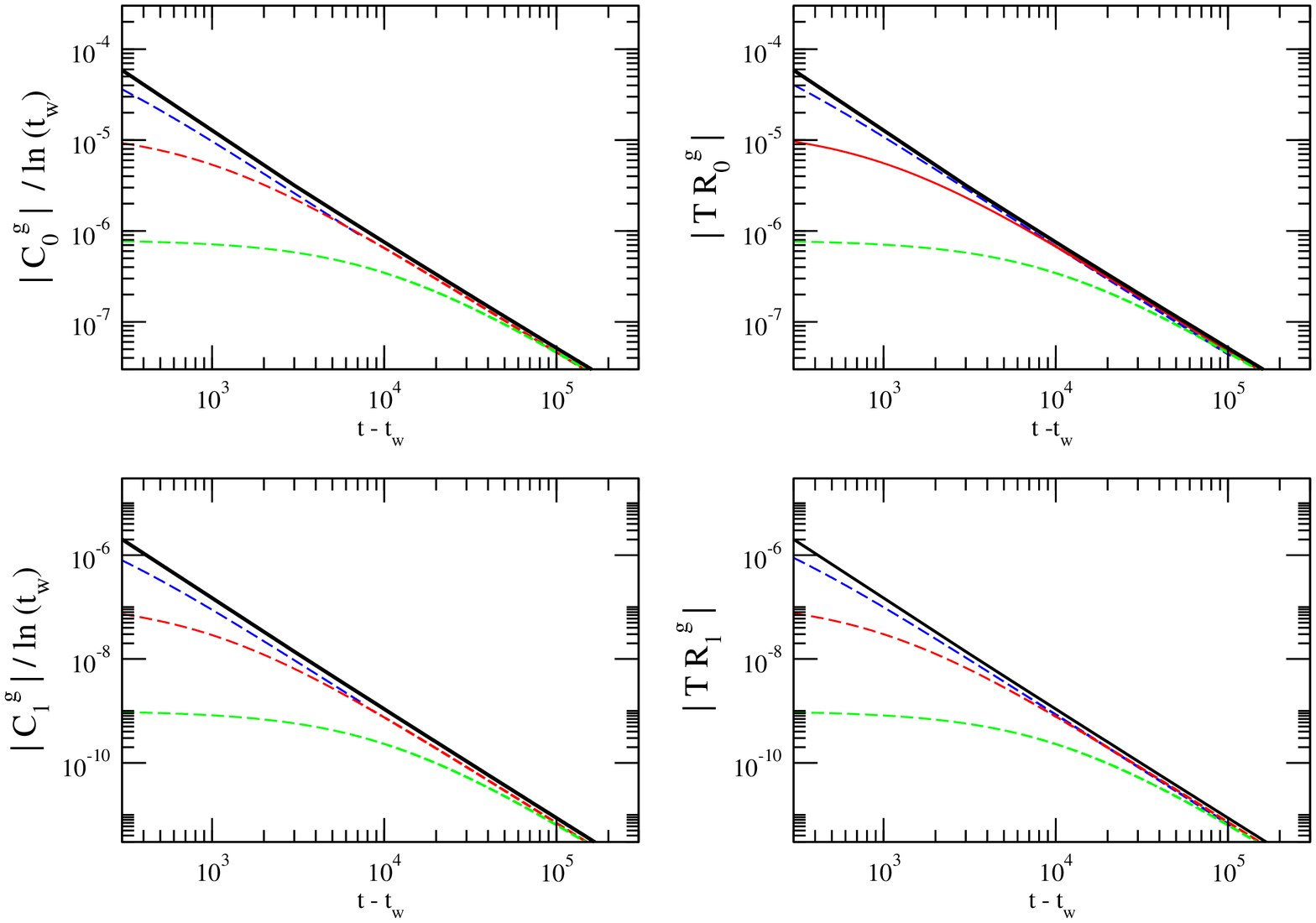}
\vskip 0.1in \caption{(Color online) Left column: scaling of the
global correlations $C^g_0$ and $C^g_1$ divided by $\ln(t_w)$. Right
column: scaling of the global correlations $R^g_0$ and $R^g_1$
multiplied by $T$. The continuous lines are the assymptotic scalings
(\ref{eq:c_sca}) and (\ref{eq:r0_sca}), (\ref{eq:r1_sca}) while the
discontinuous lines correspond, from top to bottom, to $t_w$= 100,
1000, 10000 respectively. All quantities are dimensionless and the
time is measured in Monte Carlo sweeps.\label{fig:scalings}}
\end{center}
\end{figure}

In the right column of figure \ref{fig:scalings} we numerically
confirm the scalings (\ref{eq:r1_sca}) and (\ref{eq:r0_sca}) for
different values of $t_w$. Due to the fact that the dynamical
equations for the correlations are formally identical to those for
the response functions one finds:

\beqna C^g_0(\tau,t_w)& = & -\frac{\ln(t_w)}{\tau\ln^2\tau} + {\cal{O}}\left(\frac{1}{\tau^2\ln^3\tau}\right)\nn \\
C^g_1(\tau,t_w) & = & \frac{\ln(t_w)}{\tau^2\ln \tau} +
{\cal{O}}\left(\frac{1}{\tau^2\ln^2\tau}\right)~~,
\label{eq:c_sca}\eeqna where the dependence on $t_w$ has been
inferred from the decay of the global correlations at equal times
(eq.(\ref{eq:correqtimes})).

These scalings are again confirmed numerically and are shown in the
left column of figure \ref{fig:scalings}. With these scalings we
recover the following FDRs:

\be X^g_0(t_w) = X^g_1(t_w)  \simeq -t_w ~~,\ee
in agreement with our numerical findings. A similar analysis can be
done for $k>1$.

\section{CONCLUSIONS}
\label{sec:conclusions}

In this paper we have solved the relaxation of the correlations and
response functions in the BG model for a general dynamic rule (provided that
it satisfies local detailed balance). We have studied (by means of
numerical integration and analytic asymptotic expansions) the
behavior of effective temperatures and FDRs in the glassy regime.

We  have found that both the correlation and the response functions show
two characteristic time-scales: a first $\beta$-relaxation for short
times characterized by an equilibrium FDR, $X(t,t_w)=1$, and a
second $\alpha$-relaxation at long times with a non-trivial value of
the FDR. This is a very common feature of structural glasses and
other glassy systems. Moreover, we have found that both the
correlations and responses display simple aging.

In this paper we have analyzed the resulting FDRs obtained from
local and global perturbations. The interesting conclusion is that
the local FDRs depend on both $t$ and $t_w$ while the global FDRs
only depends on $t_w$. Moreover, global FDRs are independent of the
observable in contrast with the local ones.

More interesting is the fact that this observable-independent value
of the global FDR is negative and diverges with the waiting time as:
\be X^{g}(t_w)\approx - t_w ~~~.\ee This result points in the same
direction as recent studies on kinetically constrained models
\cite{sollich_negative,sollich_negative2} which also found negative
FDRs. In these studies, the negative character of the FRDs was
associated to activation effects in the dynamics. In the present
case, we have found negative FDRs in the sole presence of entropic
barriers for the BG.

It is worth emphasizing that negative FDRs are related to
non-monotonic response functions. In the glassy literature,
non-monotonic responses are associated with \textit{non-neutral}
observable quantities~\cite{bgreviewcris}. The non-neutrality
property of these observables emerges as a consequence of the
dynamic coupling between the external field and the energy of the
system leading to negative effective temperatures.

In this paper we showed that an observable-independent FDR can be
properly defined by studying global observables. However, we found a
unique negative FDR due to the non-neutrality of the observables
under scrutinity. Therefore, the \textit{neutrality} of an
observable seems to be a key aspect in order to define
nonequilibrium effective temperatures.

How much current results would change if the
    perturbation $h \delta_{n,k}$ acts along an arbitrary direction
  $k>1$? We do not expect big qualitative changes in our results depending
  on the "orientation" of the
field provided that $k$ is finite (and $k<<N$). Arbitrary values of
$k$ will result in a bottleneck effect similar to that observed in
the current study for $k=1$. However, for $k/N$ finite the
bottleneck effect will be substantially different because the energy
will not be able to reach the asymptotic low energy regime $E\to
-1+\frac{1}{\ln(t)}~.$

Finally, it would be
  extremely helpful to find a microcanonical derivation or a
  phenomenological argument for reproducing the asymptotic behavior of
  the effective temperature when perturbing along arbitrary
    observables $P_k$. This could be done either by a closure of the
    dynamical equations by using a partial equilibration hypothesis, or
    by exact computation of the appropriate configurational entropy in the
    off-equilibrium regime. Such arguments
  would greatly facilitate the computation of effective temperatures without
  having to solve the full set of dynamical equations for correlations
  and responses.

\begin{center}
\textbf{Acknowledgments}
\end{center}

A.G. wish to thank CEAV for its support during the last stages of
this work. F.R. acknowledges support from the Spanish and Catalan
Research Councils from grants FIS2007-61433, NAN2004-9348,
SGR05-00688. I.P. acknowledges support from the Spanish and Catalan
Research Councils from grants FIS2005-01299 and SGR05-00236.

\appendix

\section{Local dynamical equations}
\label{app:local}

In the present analysis we consider a general dynamics with just one
restriction: it must obey local detailed balance. This restriction
ensures that the system converges to its equilibrium state. In fact,
we will see that this is the necessary condition for FDT to be
obeyed at equilibrium. From now on, the transition probabilities
will be expressed by $W(\Delta E)$ where $\Delta E$ is the energy
difference between the final and the initial states.

\bigskip

\subsubsection{One-time quantities.}

\bigskip

The dynamic equations for the occupation probabilities can be
computed in the same way as have been obtained for the Monte-Carlo
dynamics of this model (see Ref.\cite{bgfelix3} for details). The
general result is:

{\small{ \beqna
\frac{\partial P^1_k}{\partial t} &=& W(0)[-kP^1_k + (k+1)P^1_{k+1}-P^1_{k}+P^1_{k-1}]+\nn\\
&+& (W(0)-W(-1+h))[P^1_1(1-P_0)(\de_{k,1}-\de_{k,0})+P^1_1P_1(\de_{k,1}-\de_{k,2})]+\nn\\
&+& (W(0)-W(h))[P^1_1P_0(\de_{k,1}-\de_{k,0})+P^1_1(1-P_1)(\de_{k,1}-\de_{k,2})]+\nn\\
&+& (W(0)-W(-h))[2P^1_2(1-P_0)(\de_{k,2}-\de_{k,1})+P^1_0P_1(\de_{k,0}-\de_{k,1})]+\nn\\
&+& (W(0)-W(1-h))[2P^1_2P_0(\de_{k,2}-\de_{k,1})+P^1_0(1-P_1)(\de_{k,0}-\de_{k,1})]+\nn\\
&+& (W(0)-W(1))[P_0(kP^1_k-(k+1)P^1_{k+1})+P^1_1P_0(\de_{k,0}-\de_{k,1})+2P^1_2P_0(\de_{k,1}-\de_{k,2})]+\nn\\
&+&
(W(0)-W(-1))[P_1(P^1_k-P^1_{k-1})+P^1_0P_1(\de_{k,1}-\de_{k,0})+P^1_1P_1(\de_{k,2}-\de_{k,1})]~~.\nn\\
\label{eq:appa_1}\eeqna}}

The quantities $P_k(t)$ are the occupation probabilities, while the
quantities $P^1_k(t)$ are the average occupation probabilities
restricted to the box number one, which is the box affected by the
external field. As a particular case we can get the dynamic
evolution for the occupation probabilities for any box at zero
field:

\beqna \frac{\partial P_k}{\partial t}&=& W(0)[-kP_k +
(k+1)P_{k+1}-P_{k}+P_{k-1}]+\nn\\ &+&
(W(0)-W(-1))[P_1(P_k-P_{k-1}+\de_{k,1}-\de_{k,0})]+\nn\\ &+&
(W(0)-W(1))[P_0(kP_k-(k+1)P_{k+1}+\de_{k,0}-\de_{k,1}]~~.
\label{eq:pk_local}\eeqna

These equations cannot be solved exactly (although an analytic
treatment has been done in the asymptotic regime \cite{bgfelix3})
but can be integrated numerically to give the full solution. More
significative, these equations are the first step in order to
compute the dynamical evolution of two-time quantities such as the
autocorrelation functions and the local response functions.

\bigskip

\subsubsection{Local correlations and response functions}

\bigskip

As a consequence of the local character of the external field, we
have to deal with the corresponding local response functions and the
box-box autocorrelation functions. These autocorrelation functions
are defined as follows:

\be
C^{loc}_k(t,t_w)=\frac{1}{N}<\sum_{r}\de_{n_r(t),k}\de_{n_r(t_w),1}>~,
\ee which can be expressed in terms of the following conditional
probabilities $\nu_k(t,t_w)=P(n_r(t)=k|n_r(t_w)=0)$: \be
C^{loc}_k(t,t_w)=P_1(t_w)\nu_k(t,t_w)~. \ee

Following the same strategy as in \cite{bgfelix3} the dynamic
equations for these conditional probabilities give:

\beqna
\frac{\partial \nu_k(t,t_w)}{\partial t}&=& W(0)[-k\nu_k + (k+1)\nu_{k+1}-\nu_{k}+\nu_{k-1}]+\nn\\
&+& (W(0)-W(-1))[P_1(\nu_k-\nu_{k-1})+(\de_{k,1}-\de_{k,0})(\nu_1(1-P_0)+\nu_0P_1)]+\nn\\
&+&
(W(0)-W(1))[P_0(k\nu_k-(k+1)\nu_{k+1})+(\de_{k,0}-\de_{k,1})(\nu_0(1-P_1)+\nu_1P_0)]~.\nn\\
\label{eq:appa_2}\eeqna

The corresponding local response functions are just the variations
of the occupation probabilities for the perturbed box with the
external field:

\be R^{loc}_k=\left( \frac{\delta P_k^1(t)}{\delta h(t_w)}
\right)_{h(t_w)\rightarrow 0}~.\ee

From this expression and eq.(\ref{eq:pk_local}) we arrive at:

\beqna \frac{\partial R^{loc}_k(t,t_w)}{\partial t}&=&
W(0)[-kR^{loc}_k +
(k+1)R^{loc}_{k+1}-R^{loc}_{k}+R^{loc}_{k-1}]\nn\\
&+&(W(0)-W(-1))[P_1(R^{loc}_k-R^{loc}_{k-1})+(\de_{k,1}-\de_{k,0})(R^{loc}_1(1-P_0)+R^{loc}_0P_1)]\nn\\
&+&(W(0)-W(1))[P_0(kR^{loc}_k-(k+1)R^{loc}_{k+1})+(\de_{k,0}-\de_{k,1})(R^{loc}_0(1-P_1)+R^{loc}_1P_0)]\nn\\
&+&\de(t-t_w)S^{loc}[<P_k>] \label{eq:appa_3} \eeqna

Note that formally, the dynamic evolution for the response functions
is just the same as for the autocorrelation functions. This is a
general feature and is due to the fact that in equilibrium FDT must
be satisfied. The only difference is that in the equation for the
responses there is a delta term which fixes the value for
$R^{loc}_k(t_w,t_w)$. This term comes from the first order of the
Taylor expansion in the transition probabilities which depend on the
external field $h$. This is not an approximation because higher
order terms in Taylor's expansion vanish when we set the external
field equal to zero. The function $S^{loc}[<P_k>]$ is defined as:
\beqna S^{loc}[<P_k>]&=&\beta
e^{\beta}W(1)[P_1(1-P_0)(\de_{k,1}-\de_{k,0})+P_1^2(\de_{k,1}-\de_{k,2})]+\nn\\
&+& \beta
e^{\beta}W'(1)[P_1(1-P_0)(\de_{k,1}-\de_{k,0})+P_1^2(\de_{k,1}-\de_{k,2})]-\nn\\
&-& \beta
W(0)[2P_2(1-P_0)(\de_{k,2}-\de_{k,1})+P_1P_0(\de_{k,0}-\de_{k,1})]-\nn\\
&-&\beta W'(0)[2P_2(1-P_0)(\de_{k,2}-\de_{k,1})]+\nn\\ &+&\beta
W'(1)[2P_2P_0(\de_{k,2}-\de_{k,1})+(1-P_1)P_0(\de_{k,0}-\de_{k,1})]-\nn\\
&-&\beta
W'(0)[(1-P_1)P_0(\de_{k,0}-\de_{k,1})]~,\label{eq:appa_delta} \eeqna
where W' denotes the derivative of the transition probability with
respect to $\Delta E$. Finally, we must stress that in this equation
we have already supposed that our dynamics verifies local detailed
balance. If W' is discontinuous we should have to consider two
possible response functions depending on the chosen value for
W'~\cite{bggodreche3}.

\bigskip

\section{Global dynamical equations}
\label{app:global}

As we have made in the case of a local perturbation, we consider a
general dynamics with the only condition that it must obey detailed
balance. As before, this is the unique ingredient we need to ensure
that equilibrium is reached at long enough times.

\bigskip

\subsubsection{One-time quantities.}

\bigskip

By considering all the possible elementary moves, we get the
following dynamical equations for the occupation probabilities:

\beqna
\frac{\partial P_k}{\partial t} &=& W(0)[-kP_k + (k+1)P_{k+1}-P_{k}+P_{k-1}]+\nn\\
&+& (W(0)-W(h-1))[P_1(1-P_1)(\de_{k,1}-\de_{k,0})+P_kP_1(1-\de_{k,1})-P_{k-1}P_1(1-\de_{k,2})]+\nn\\
&+& (W(0)-W(2h-1))[P_1^2(2\de_{k,1}-\de_{k,0}-\de_{k,2})]+\nn\\
&+& (W(0)-W(-h))[2P_2(1-P_0)(\de_{k,2}-\de_{k,1})+2P_kP_2(1-\de_{k,0})-2P_{k-1}P_2(1-\de_{k,1})]\nn\\
&+& (W(0)-W(1-2h))[-2P_0P_2(2\de_{k,1}-\de_{k,0}-\de_{k,2})]+\nn\\
&+& (W(0)-W(1-h))[P_0(kP_k-(k+1)P_{k+1}+\de_{k,0}-\de_{k,1}+2P_2(2\de_{k,1}-\de_{k,0}-\de_{k,2}))]\nn\\
&+& (W(0)-W(h))[P_1(kP_k-(k+1)P_{k+1}+\de_{k,1}-\de_{k,2})+P_1^2(-2\de_{k,1}+\de_{k,0}+\de_{k,2})]~.\nn\\
\label{eq:appb_1}\eeqna

Note that now, due to the global character of the field, we have to
consider only the occupation probabilities averaged over the whole
system. As we expect, at zero field we recover the same equations as
in the local case (which are the extension of the equations obtained
for Monte Carlo dynamics \cite{bgfelix2}):

\beqna
\frac{\partial P_k}{\partial t} &=& W(0)[-kP_k + (k+1)P_{k+1}-P_{k}+P_{k-1}]+\nn\\
&+& (W(0)-W(-1))[P_1(\de_{k,1}-\de_{k,0}+P_k-P_{k-1})]+\nn\\
&+& (W(0)-W(1))[P_0(kP_k-(k+1)P_{k+1}+\de_{k,0}-\de_{k,1})] ~.
\eeqna

These equations are the first step in order to compute the dynamical
equations for the correlation and response functions and give the
evolution of all the possible observable physical quantities of this
model. Moreover, these equations are the base of the more complex
computations of the dynamical evolution of the two-time correlation
and response functions.

\bigskip

\subsubsection{Global correlations and response functions}

\bigskip

Due to the extensive nature of the perturbation the correlation
functions related with the responses are the connected ones. So, let
us introduce the deviation of the instantaneous values of the
occupations from their average value at each time:

\be \gamma_k(t)=\frac{1}{N} \sum_r \de_{n_r,k} - P_k(t)~, \ee

These quantities will give us insight on the fluctuations of the
occupation numbers (i.e the correlations). The dynamical evolution of these quantities is:

\beqna
\frac{\partial \gamma_k}{\partial t} &=& W(0)[-k\gamma_k + (k+1)\gamma_{k+1}-\gamma_{k}+\gamma_{k-1}]+\nn\\
&+& (W(0)-W(-1))[\gamma_1(\de_{k,1}-\de_{k,0}+P_k-P_{k-1})+P_1(\gamma_k-\gamma_{k-1})]+\nn\\
&+&
(W(0)-W(1))[\gamma_0(kP_k-(k+1)P_{k+1}+\de_{k,0}-\de_{k,1})+P_0(k\gamma_k-(k+1)\gamma_{k+1})].\nn\\
 \eeqna

In this equation we have considered that the quantities $\gamma_k$
are of order $1/N$, so we have neglected the quadratic terms
$\gamma_k\gamma_l$ in these equations because they vanish in the
thermodynamic limit. The global connected correlation function will
be:

\be C^g_k(t,t_w)=<\gamma_k(t)\gamma_1(t_w)>~. \ee

The equations of motion for these correlations are easy to compute
from these equations and give:

\beqna
\frac{\partial C^g_k(t,t_w)}{\partial t} &=& W(0)[-kC^g_k + (k+1)C^g_{k+1}-C^g_{k}+C^g_{k-1}]+\nn\\
&+& (W(0)-W(-1))[C^g_1(\de_{k,1}-\de_{k,0}+P_k-P_{k-1})+P_1(C^g_k-C^g_{k-1})]+\nn\\
&+&
(W(0)-W(1))[C^g_0(kP_k-(k+1)P_{k+1}+\de_{k,0}-\de_{k,1})+P_0(kC^g_k-(k+1)C^g_{k+1})].\nn\\
\label{eq:appb_2}\eeqna

Now we define the global response function (which is related to the
experimental susceptibility) as the response of the probabilities to
the extensive perturbation coupled to $P_1$:

\be R^g_k(t,t_w)=\left( \frac{\delta P_k(t)}{\delta h(t_w)}
\right)_{h(t_w)\rightarrow 0}~. \ee

From the equations in a field and expanding to first order in $h$ we
have for the response functions:

\beqna \frac{\partial R^g_k(t,t_w)}{\partial t} &=& W(0)[-kR^g_k +
(k+1)R^g_{k+1}-R^g_{k}+R^g_{k-1}]+\nn\\ &+&
(W(0)-W(-1))[R^g_1(\de_{k,1}-\de_{k,0}+P_k-P_{k-1})+P_1(R^g_k-R^g_{k-1})]+\nn\\
&+&
(W(0)-W(1))[R^g_0(kP_k-(k+1)P_{k+1}+\de_{k,0}-\de_{k,1})+P_0(kR^g_k-(k+1)R^g_{k+1})]+\nn\\
&+& \de(t-t_w) S^g[<P_k>]~,\label{eq:appb_3}\eeqna where we have
defined the function $S^g[<P_k>]$ which depends only on one time and
gives the initial value for the responses. Similar computations as
we have done for the local case lead to:

\beqna
S^g[<P_k>]&=&\beta e^{\beta}W(1)[P_1(1-P_1)(\de_{k,1}-\de_{k,0})+P_kP_1(1-\de_{k,1})+P_{k-1}P_1(1-\de_{k,2})]+\nn\\
&+& \beta e^{\beta}W'(1)[P_1(1-P_1)(\de_{k,1}-\de_{k,0})+P_kP_1(1-\de_{k,1})+P_{k-1}P_1(1-\de_{k,2})]+\nn\\
&+& 2\beta e^{\beta}(W(1)+W'(1))[P_1^2(2\de_{k,1}-\de_{k,0}-\de_{k,2})]-\nn\\
&-& \beta W(0)[2P_2(1-P_0)(\de_{k,2}-\de_{k,1})+2P_kP_2(1-\de_{k,0})-2P_{k-1}P_2(1-\de_{k,1})]-\nn\\
&-& \beta W'(0)[2P_2(1-P_0)(\de_{k,2}-\de_{k,1})+2P_kP_2(1-\de_{k,0})-2P_{k-1}P_2(1-\de_{k,1})]+\nn\\
&+& 2\beta W'(1)[-2P_0P_2(2\de_{k,1}-\de_{k,0}-\de_{k,2})]+\nn\\
&+& \beta W'(1)[P_0(kP_k-(k+1)P_{k+1}+\de_{k,0}-\de_{k,1})+2P_0P_2(2\de_{k,1}-\de_{k,0}-\de_{k,2})]-\nn\\
&-&
\beta W'(0)[P_1(kP_k-(k+1)P_{k+1}+\de_{k,1}-\de_{k,2})+P_1^2(-2\de_{k,1}+\de_{k,0}+\de_{k,2})].\nn\\
\label{eq:appb_delta}\eeqna

As before $W'(\Delta E)$ is the derivative of the transition
probability with respect to $\Delta E$ evaluated at $\Delta E=0,~1$.


\begin{thebibliography}{08}

\bibitem{jou} J. Casas-v\'azquez and D. Jou. Temperature in nonequilibrium states:
a review of open problems and current proposals. {\it Rep. Prog.
Phys. \textbf{66} 1937-2023, (2003).}

\bibitem{young} Sping Glasses and Random Fields, edited by A.P.
Young (World Scientific, Singapore, 1998).

\bibitem{kubo} R.Kubo, \textit{Repr. Progr. Phys \textbf{29} 255,
1966.}; R. Kubo, M. Toda and N. Hashitsume, \textit{Statistical
Physics II} (2nd edn. Springer Verlag, Berlin, 1991).

\bibitem{kurchan} L.F. Cugliandolo, J. Kurchan, and L. Peliti.
Energy flow, partial equilibration and effective temperatures in
systems with slow dynamics.
\newblock \textit {Phys. Rev. E, {\bf 55(4)}:3898--3914, (1997)}.

\bibitem{bgreviewcris} A. Crisanti and F. Ritort. Violations of
the fluctuation-dissipation theorem in glassy systems: basic notions
and the numerical evidence. \textit{J. Phys. A (Math. Gen)
\textbf{36} R181, 2003.}

\bibitem{bggarrahan2} P. Mayer, L. Berthier, J.P. Garrahan and P.
Sollich. Fluctuation-dissipation relations in the nonequilibrium
critical dynamics of Ising models.\textit{Physical Review E,
\textbf{68} 016116, 2003.}

\bibitem{sollich_negative} P. Mayer, S. L\'eonard, L. Berthier, J.P.
Garrahan and P. Sollich. Activated aging dynamics and negative
fluctuation-dissipation ratios. {\it Phys. Rev. Lett. 96, 030602
(2006).}

\bibitem{sollich_negative2} S. L\'eonard, P. Mayer, P. Sollich, L. Berthier and J.P.
Garrahan. Nonequilibrium dynamics of spin facilitated glass models.
{\it J. Stat. Mech. P07017 (2007)}.

\bibitem{bgfelix1} F. Ritort. Glassiness in a model without energy
barriers. \textit{Physical Review Letters, \textbf{75} 1190,
1995.}

\bibitem{bgfelix2} S. Franz and F. Ritort. Dynamical solution of a model
without energy barriers.\textit{Europhysics Letters, \textbf{31}
507, 1995.}

\bibitem{bgfelix3} S. Franz and F. Ritort. Glassy mean-field dynamics of
the backgammon model. \textit{Journal of Statistical Physics,
\textbf{85} 131, 1996}

\bibitem{bgfelix4} S. Franz and F. Ritort. Relaxation processes and
entropic traps in the backgammon model. \textit{Journal of Physics
A, \textbf{30} L359, 1997.}

\bibitem{bggodreche1} C. Godreche, J.P. Bouchaud and M. Mezard. Entropy
barriers and slow relaxation in some random walk models.
\textit{Journal of Physics A, \textbf{28} L603, 1995.}

\bibitem{bggodreche2} C. Godreche and J.M. Luck. Long-time regime and
scaling of correlations in a simple model with glassy behavior.
\textit{Journal of Physics A, \textbf{29} 1915, 1996.}

\bibitem{bggodreche3} C. Godreche and J.M. Luck. Correlation and response
in the Backgammon model: the Ehrenfest legacy. \textit{Journal of
Physics A, \textbf{32} 6033, 1999.}

\bibitem{bgprados} A. Prados, J.J. Brey and B. Sanchez-Rey. Glassy
behavior in a simple model with entropy barriers. \textit{Physical
Review B, \textbf{55} 6343, 1997.}



\bibitem{bgreviewsoll} P. Sollich and F. Ritort. Glassy dynamics of
kinetically constrained models. \textit{Advances in Physics,
\textbf{52} 219, 2003.}

\bibitem{bgehrenfest} P. Ehrenfest and T. Ehrenfest. \textit{The
conceptual foundations of the statistical approach to mechanics.}
Dover, 1990.


\bibitem{bgehrenfest2} P. Ehrenfest and T. Ehrenfest. \textit{Phys. Zeit.
\textbf{8} 311, 1907.}

\bibitem{bgluckurn} C. Godreche and J.M. Luck. Nonequilibrium dynamics of
urn models. \textit{J. Phys. Cond. Matt., \textbf{14} 1601, 2002.}

\bibitem{bookleuzzi} L. Leuzzi and Th. M Nieuwenhuizen. Thermodynamics of the Glassy
State. \textit{Taylor and Francis (2007).}

\bibitem{bgleuzzi} L. Leuzzi and F. Ritort. Disordered Backgammon
Model. \textit{Physical Review E, \textbf{65} 56125, 2002.}

\bibitem{bgferro} A. Garriga, P. Sollich, I. Pagonabarraga and F.
Ritort. Universality of Fluctuation-Dissipation Relations. The
Ferromagnetic Model. \textit{Physical Review E, \textbf{72} 056114,
2005.}




%\bibitem{bg_mayer} P. Mayer and P. Sollich. General Solutions for
%Multispin Two-Time Correlation and Response Functions in the
%Glauber-Ising Chain. Preprint Cond-mat/0307214.




%\bibitem{bggarrahan} A.Buhot and J.P. Garrahan. Fluctuation-dissipation
%relations in the activated regime of simple strong-glass models. {\it
%Phys. Rev. Lett. 88, 225702 (2002).}


%\bibitem{sollich_esf} A. Annibale and P. Sollich. Spin, bond and global
%fluctuation dissipation relations in the nonequilibrium spherical
%ferromagnet. {\it J. Phys. A: Math. Gen. 39 2853-2907 (2006).}




\end{thebibliography}
\end{document}